\documentclass[a4paper,fleqn,usenatbib]{mnras}

\usepackage{savesym}
\usepackage{amsmath}
\savesymbol{iint}
\usepackage{txfonts}
\restoresymbol{TXF}{iint}

\usepackage[T1]{fontenc}
\usepackage{ae,aecompl}

\usepackage{graphicx}	
\usepackage{amsmath}	
\usepackage{amssymb}	
\usepackage{natbib}
\usepackage{subcaption,tikz}
\tikzset{boximg/.style={remember picture,red,thick,draw,inner sep=0pt,outer sep=0pt}}

\title[Hypervelocity runaways from the LMC]{Hypervelocity runaways from the Large Magellanic Cloud}

\author[D. Boubert et al.]{
D. Boubert,$^{1}$\thanks{E-mail: d.boubert,nwe,derkal,rgi@ast.cam.ac.uk}
D. Erkal,$^{1}$
N. W. Evans$^{1}$
and R. G. Izzard$^{1}$
\\
$^{1}$Institute of Astronomy, University of Cambridge, Madingley Road, Cambridge CB3 0HA, UK\\
}

\date{Accepted XXX. Received YYY; in original form ZZZ}

\pubyear{2017}

\begin{document}
\label{firstpage}
\pagerange{\pageref{firstpage}--\pageref{lastpage}}
\maketitle

\begin{abstract}
We explore the possibility that the observed population of Galactic
hypervelocity stars (HVSs) originate as runaway stars from the Large
Magellanic Cloud (LMC). Pairing a binary evolution code with an N-body
simulation of the interaction of the LMC with the Milky Way, we
predict the spatial distribution and kinematics of an LMC
runaway population. We find that runaway stars from the LMC can
contribute Galactic HVSs at a rate of $3 \times
10^{-6}\;\mathrm{yr}^{-1}$. This is composed of stars at different
points of stellar evolution, ranging from the main-sequence to those
at the tip of the asymptotic giant branch. We find that the known
B-type HVSs have kinematics which are consistent with an LMC
origin. There is an additional population of hypervelocity white
dwarfs whose progenitors were massive runaway stars. Runaways which
are even more massive will themselves go supernova, producing a
remnant whose velocity will be modulated by a supernova kick. This
latter scenario has some exotic consequences, such as pulsars and
supernovae far from star-forming regions, and a small rate of
microlensing from compact sources around the halo of the LMC.
\end{abstract}




\section{Introduction}
\label{sec:intro}
In this decade of large and precise kinematic datasets, it is tempting
to go hunting for outliers. These range from the unusual
$30\text{--}40 \;\mathrm{km}\;\mathrm{s}^{-1}$ OB-runaways, first
explained by \cite{blaauw_origin_1961} as the runaway former
companions of supernova progenitors, to the extreme $>500\;\mathrm{km}\;\mathrm{s}^{-1}$ hypervelocity stars (HVSs), which
are unbound from the Milky Way (MW). The latter are suspected to have
been accelerated by the Hills mechanism, where the tidal disruption of a binary by the massive black
hole (SMBH) Sgr A* in the Galactic Centre results in the rapid ejection of one of the stars
\citep{hills_hyper-velocity_1988}. Alternative explanations for these
stars include dynamical ejection from young clusters
\citep{perets_dynamical_2009}, extreme supernova runaway scenarios
\citep{portegies_zwart_characteristics_2000}, tidal debris from an
accreted dwarf galaxy \citep{abadi_alternative_2009} or an SMBH in the
Large Magellanic Cloud \citep{boubert_dipole_2016}.

In this work, we explore the consequences of the production of runaway
stars in the Large Magellanic Cloud (LMC). The LMC has a
star-formation rate of $0.2\;\mathrm{M}_{\odot}\;\mathrm{yr}^{-1}$
\citep{harris_star_2009} and an orbital velocity of $378 \;
\mathrm{km} \; \mathrm{s}^{-1}$
\citep{van_der_marel_third-epoch_2014}, so it is plausible that
runaway stars from the LMC could contribute a meaningful proportion of
the Galactic HVSs. More massive galaxies typically have more of
everything (e.g. globular clusters and supernova), but this does not necessarily
include having more unbound, escaping stars.  This is because lowering the mass of the
galaxy lowers the required escape velocity. As most stellar processes
which produce high velocity stars have a steeply decreasing
distribution with increasing velocity, it follows that decreasing the
mass of a galaxy can result in more escaping stars. The velocities
produced by these processes are set by stellar properties, and these
are only weakly dependent on the host galaxy. Some other possible
origins of unbound LMC stars include stripping from the LMC by a
previous passage of the SMC \citep{besla_origin_2013}, formation in
the gas of the leading arm of the LMC
\citep{casetti-dinescu_recent_2014}, and ejection from the LMC
either by a putative SMBH in the centre of the LMC
\citep{edelmann_he_2005,boubert_dipole_2016} or dynamical interactions
in a stellar cluster, possibly involving an intermediate mass black hole
\citep{gualandris_hypervelocity_2007}.

The field of fast-moving stars is beset by a muddle of nomenclature,
which stems from the difference between classifying stars by how fast
they are moving or by their origin. Among HVSs this is a crucial
distinction. A star may be ejected by the Hills mechanism, but remain
bound to the galaxy. Conversely, a star may be unbound, but not
produced by the Hills mechanism. Runaway stars are usually defined as
OB stars with peculiar velocities in excess of
$40\;\mathrm{km}\;\mathrm{s}^{-1}$ \citep{blaauw_origin_1961}, with
either dynamical ejection from a young cluster or a supernova ejecting
the progenitor's companion as their origin. However, the slower
cousins of the binary supernova runaways are also termed runaways by
several authors, with increasing use of the term walkaways for those
runaways ejected slower than $10\;\mathrm{km}\;\mathrm{s}^{-1}$
\citep{de_mink_challenges_2012,de_mink_incidence_2014,lennon_gaia_2016}.
A convention sometimes used in the literature is to refer to unbound
Hills stars as hypervelocity and unbound runaway stars as hyperrunaway
\citep[eg.][]{perets_properties_2012,brown_hypervelocity_2015}. However,
this is open to the objection that it is in practice difficult to
determine the origin of the known unbound stars in the Galaxy. For
example, they may not originate in the Milky Way
\citep[e.g.,][]{boubert_dipole_2016} and -- as we show in this paper
-- they may not even originate with the Hills mechanism.  To clarify
the terminology of this paper, we exclusively use the term {\it
  runaway} to refer to stars of all velocities whose binary companion
has gone supernova and the term {\it hypervelocity} to refer to stars
of any origin which are unbound from the Milky Way. All stars emitted
from a binary tidally disrupted by a central black hole of either
galaxy are \emph{Hills stars}. To avoid the confusion of referring to
stars which escape the LMC as HVSs with respect to the LMC, we will
use the terms LMC remainers/escapers to refer to stars which are
bound/unbound to the LMC.


In Section \ref{sec:method}, we describe the method we use to generate
runaways and then follow their stellar evolution and orbit in an
LMC-MW potential. There are many observables associated with runaway
stars which escape the LMC and we discuss these in Section
\ref{sec:prop}. Our conclusions in Section \ref{sec:conclusion} are
that runaway stars escaping the LMC must contribute to the Milky Way
hypervelocity star population, but that the stellar types and
distribution of these hypervelocity runaways are dependent on the
assumed binary evolution model.

\section{LMC runaway ejection model}
\label{sec:method}

There are several ingredients required for a model of the ejection of
runaway stars from the LMC. Assuming a metallicity and star formation
history for the LMC, we evolve a synthetic population of single and
binary stars and identify the runaway stars. The runaways are then
initialised in the LMC disk and their subsequent orbits integrated
through an evolving N-body potential of the LMC and the Galaxy. The
outcomes of the stellar evolution of these runaway stars and their
kinematics are then transformed into observable properties.

\subsection{Star Formation History of the LMC}
\label{sec:sfh}

Our method requires knowledge of the time dependent star-formation
rate (SFR) and metallicity of the LMC. \cite{harris_star_2009} found
that the star formation rate of the LMC over the past 5 Gyr has been
constant at $0.2\;\mathrm{M}_{\odot}\;\mathrm{yr}^{-1}$ within a
factor of two. However, this period of constancy was preceded by a
quiescent epoch between $5$ and $12\;\mathrm{Gyr}$ ago. We thus assume
a constant star formation rate over the entire $1.97\;\mathrm{Gyr}$ we
simulate.

\cite{piatti_age-metallicity_2013} investigated the age-metallicity
relation for the LMC using photometry across 21 fields. They derived
an approximate scaling,
\begin{equation}
\label{eq:amr}
[\mathrm{Fe}/\mathrm{H}]=C+\left(\frac{\partial [\mathrm{Fe}/ \mathrm{H}]}{\partial t}\right) t + \left(\frac{\partial [\mathrm{Fe}/ \mathrm{H}]}{\partial a}\right) a,
\end{equation}
with $C=-0.55\pm0.02\;\mathrm{dex}$, $\partial [\mathrm{Fe}/
  \mathrm{H}]/\partial t = -0.047\pm0.003\;\mathrm{dex} \;
\mathrm{Gyr}^{-1}$ and $\partial [\mathrm{Fe}/ \mathrm{H}]/\partial a
= -0.007 \pm 0.006\;\mathrm{dex} \; \mathrm{degree}^{-1}$, where $a$
is the de-projected angular distance from the centre of the LMC. The
dependency on the angular distance is argued by
\cite{piatti_age-metallicity_2013} to be negligible, because under the
assumption of an LMC distance of $50\;\mathrm{kpc}$ it corresponds to
a gradient of $-0.01\pm0.01\;\mathrm{dex}\;\mathrm{kpc}^{-1}$. Thus,
we assume a constant metallicity throughout the LMC star-forming
regions. Over the $1.97\;\mathrm{Gyr}$ of our simulations, even the
temporal gradient is mostly negligible, producing a change in
$[\mathrm{Fe}/\mathrm{H}]$ of $-0.093\pm0.006\;\mathrm{dex}$. Thus, we
form stars at a constant metallicity of $Z=0.008$.

Most stars form in clusters \citep{lada_embedded_2003}, but this does
not mean that star formation in the LMC is clumpy. The currently
most prominent star-forming region in the LMC is 30 Doradus, also
known as the Tarantula Nebula. \cite{de_marchi_star_2011} found the
star formation rate to be of the order
$200\;\mathrm{M}_{\odot}\;\mathrm{Myr}^{-1}$ over at least the last
$30\;\mathrm{Myr}$ for objects in the mass range
$0.5-4.0\;\mathrm{M}_{\odot}$. This is consistent with the more recent
work of \citet{cignoni_hubble_2015} who, as part of the Hubble
Tarantula Treasury Project, found that the star formation rate (SFR)
in 30 Doradus has exceeded the average LMC SFR for the last
$20\;\mathrm{Myr}$. While 30 Doradus is one of the most active
star formation regions in the Local Group, comparing its rate
$200\;\mathrm{M}_{\odot}\;\mathrm{Myr}^{-1}$ to the rate for the
entire LMC $0.2\;\mathrm{M}_{\odot}\;\mathrm{yr}^{-1}$ reveals that 30 Doradus makes up
only $0.2\%$ of the recent star formation activity of the LMC. We are
thus well justified in forming stars directly proportional to the
density of the assumed LMC disk potential and neglecting any
inhomogeneities due to star forming clusters. We note that if this
assumption does break down, it would reveal itself as a skewed density
distribution of the ejected stars on the sky. This is because the
location from which runaway stars are ejected is encoded in the
velocity of those runaways through the contribution of the LMC disk
rotation at the location of ejection.

\subsection{Binary Evolution}
\label{sec:grid}

A standard prescription for the distribution of runaway star ejection
velocities $v_{\mathrm{ej}}$ is an exponential law in the form
$\exp(-v_{\mathrm{ej}}/v_{\mathrm{s}})$, where $v_{\mathrm{s}}\approx
150\;\mathrm{km}\;\mathrm{s}^{-1}$ is a characteristic velocity which
sets the width of the distribution (used by
\citealt{bromley_runaway_2009} and \citealt{kenyon_predicted_2014} who
matched to binary star simulations of
\citealt{portegies_zwart_characteristics_2000}). However, this
velocity distribution is simplistic because the highest ejection
velocities require close binaries. Close binaries interact, making the
ejection velocities of runaways a sensitive function of the binary
initial conditions. Given that the magnitude and colour of stars can 
be thought of broadly as a proxy for their mass and that one of the most 
important parameters in binary interaction is the ratio of masses $q$, 
the colour and ejection velocity of a runaway star must be interdependent.

We form stars in bursts every $1\;\mathrm{Myr}$. This is driven by a
computational consideration to allow for a simple implementation of
star formation in which we sample single and binary stars from
analytic distributions until we have formed the required mass of
stars. A SFR of $0.2\;\mathrm{M}_{\odot}\;\mathrm{yr}^{-1}$ means we
are thus forming starburts with a total mass of $2\times10^5
\;\mathrm{M}_{\odot}$. Only a small fraction of this mass is used to form
runaways. Our model population consists of both single stars and binaries,
but no higher-order multiples are considered. To generate the
population, we sample in the primary mass and, for binaries, in the
mass ratio and initial period. We sample systems one-by-one until we
have formed the required total mass of stars in a timestep.

We first sample the primary mass of each system from the
\mbox{\cite{kroupa_variation_2001}} IMF,
\begin{equation}
N(M_1)\propto
\begin{cases}
M_1^{-0.3}, & \mathrm{if}\ 0.01<M_1/\mathrm{M}_{\odot}<0.08, \\
M_1^{-1.3}, & \mathrm{if}\ 0.08<M_1/\mathrm{M}_{\odot}<0.5, \\
M_1^{-2.3}, & \mathrm{if}\ 0.5<M_1/\mathrm{M}_{\odot}<80.0, \\
0, & \mathrm{otherwise.}
\end{cases}
\end{equation}
We calculate the binary fraction as a function of primary
mass. \cite{arenou_simulated_2010} provides an analytic empirical fit
to the observed binary fraction of various stellar masses,
\begin{equation}
F_{\mathrm{bin}}(M_1)=0.8388\tanh(0.079+0.688M_1).
\end{equation}
We validate this formula by comparing to the data of
\citet{raghavan_survey_2010}, who provide the binary fraction as a
function of spectral type. The binary fraction has only been well
studied in the Milky Way and it is possible that the lower
metallicity stars in the LMC could exhibit a different dependency on
the primary mass. Our results turn out not to be overly dependent on
the exact form assumed for the dependency of the binary fraction on
the primary mass. This is because in our grid of evolved binary
systems most runaways come from high-mass systems in which the binary
fraction is close to unity in all prescriptions. We assume a flat
mass-ratio distribution for each system over the range
$0.1\;\mathrm{M}_{\odot}/M_1<q<1$. The period distribution is taken
from \cite{duquennoy_multiplicity_1991} and is a normal distribution
in $\log_{10}(P/\mathrm{days})$ with a mean of 4.8 and a standard
deviation of 2.3, truncated to lie between -2.0 and 12.0. The observed
period distribution of OB-type stars is closer to being log-uniform
\citep{opik_statistical_1924,sana_binary_2012}, however the error
incurred by this choice is subdominant to the uncertainty in the
outcome of the common-envelope phase.


%
We model the properties of stars ejected from binary systems in which
one component goes supernova using the {\sc binary\_c}
population-nucleosynthesis framework
\citep{izzard_new_2004,izzard_population_2006,izzard_population_2009}.
      {\sc binary\_c} is based on the binary-star evolution ({\sc
        bse}) algorithm of \citet{hurley_evolution_2002} updated to
      include nucleosynthesis, wind-Roche-lobe-overflow
      \citep{abate_wind_2013,abate_modelling_2015}, stellar
      rotation \citep{de_mink_rotation_2013}, accurate stellar
      lifetimes of massive stars \citep{schneider_ages_2014},
      dynamical effects from asymmetric supernovae
      \citep{tauris_runaway_1998}, an improved algorithm describing
      the rate of Roche-lobe overflow \citep{claeys_theoretical_2014}
      and core-collapse supernovae
      \citep{zapartas_delay-time_2017}. In particular, we take our
      black hole remnant masses from \citet{spera_mass_2015}, use a
      fit to the simulations of \citet{liu_interaction_2015} to
      determine the impulse imparted by the supernova ejecta on the
      companion and assume that the natal kick on the comapct remnants of Type II supernovae is Maxwellian \citep{hansen_pulsar_1997}, all of which were options previously implemented in
      {\sc binary\_c}. We use version 2.0pre22, SVN 4585. Grids of
      stars are modelled using the {\sc binary\_grid2} module to
      explore the single-star parameter space as a function of stellar
      mass $M$, and the binary-star parameter space in primary mass
      $M_{1}$, secondary mass $M_{2}$ and orbital period $P$.

The initial conditions of the binaries sampled are compared to a
binary grid and we identify all runaways which are formed by this
population. We pre-compute this binary grid of 8,000,000 binaries with
primary mass $M_1$, mass ratio $q$ and period $P$ having the ranges,
\begin{align}
8.0 \leq M_1 / \mathrm{M}_{\odot} &\leq 80.0,\nonumber\\
0.1 \; \mathrm{M}_{\odot}/M_1  \leq q &\leq 1,\\
-2.0 \leq \log_{10} (P/\mathrm{days}) &\leq 12.0.\nonumber
\end{align}

\begin{figure}
\includegraphics[scale=0.44,trim = 3.5mm 0mm 0mm 0mm, clip]{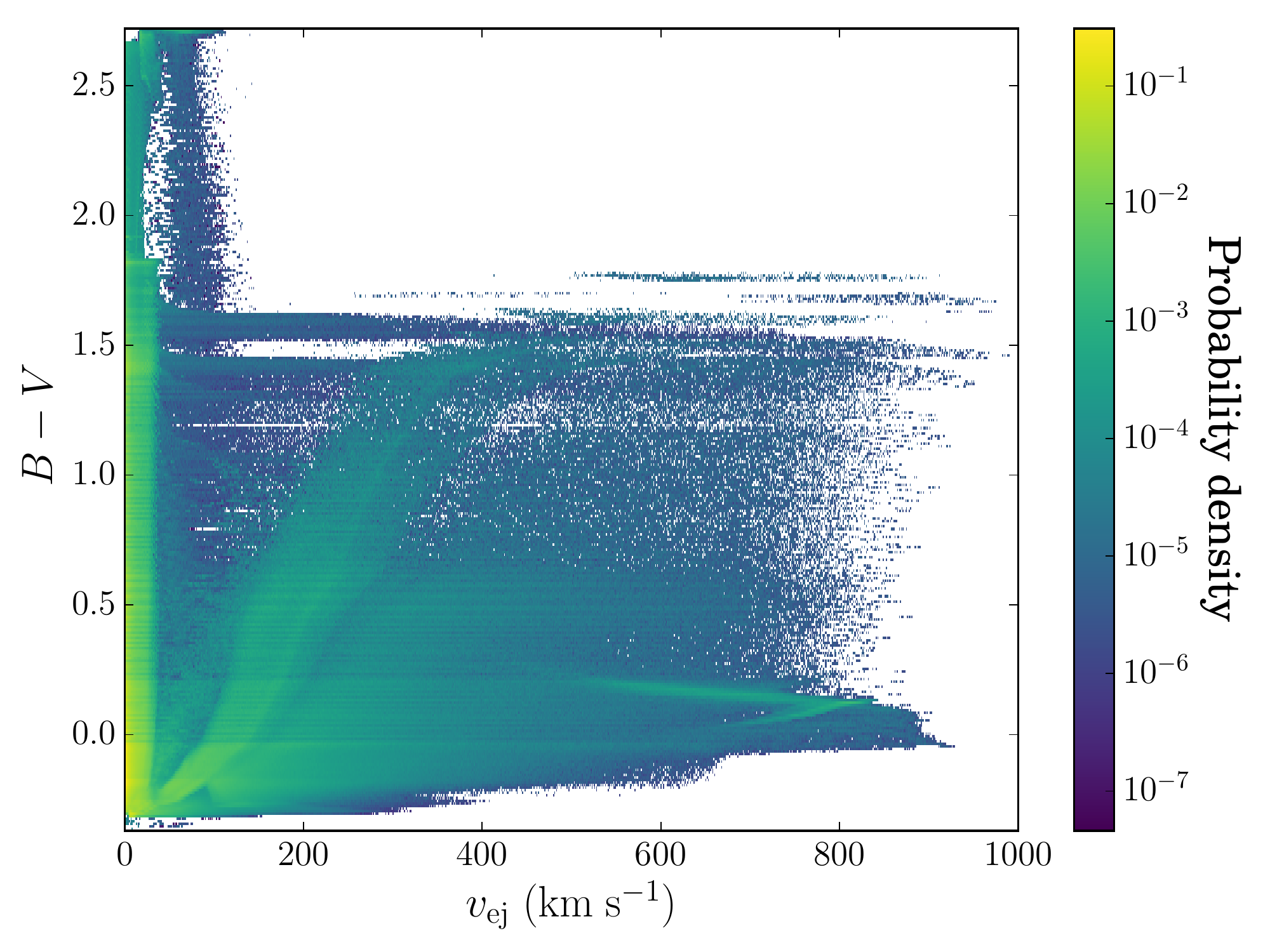}

\caption{Probability density distribution in velocity-colour space at the time of ejection from the progenitor binary of
  the runaways produced by our binary evolution grid assuming LMC
  metallicity $Z=0.008$ and common-envelope ejection efficiency
  $\alpha_{\mathrm{CE}}=1.0$.}
\label{fig:bvvej}
\end{figure}

The distribution of runaway ejection velocity $v_{\mathrm{ej}}$ and
$B-V$ colour at the time of ejection from the progenitor binary is
shown in Fig.~\ref{fig:bvvej}. A discussion of the detailed structure
in the data is elsewhere (Boubert et al, in prep.)  but the
distribution can be divided into two regions. The slower runaways with
$v_{\mathrm{ej}}<30\;\mathrm{km}\;\mathrm{s}^{-1}$ are the classical
runaways in which the progenitor binary does not interact prior to the
supernova. Conversely, runaways with
$v_{\mathrm{ej}}>30\;\mathrm{km}\;\mathrm{s}^{-1}$ are those whose
progenitor binary did interact. When the primary evolves to the giant
branch, it overflows its Roche-lobe onto the companion, provided the
companion is sufficiently close \citep{de_marco_dawes_2017}. Mass
transfer from a higher mass star to a lower mass star shrinks the
binary orbit and increases the rate of mass transfer. This process is
self-reinforcing and leads to common envelope evolution and further
shrinkage of the binary. If the common envelope is dissipated before 
the stars merge, the binary is left in a close orbit. When the 
primary does go supernova shortly afterwards the natal kick on the 
remnant may be sufficient to unbind this close binary. In
this case the rapid orbital velocity of the companion prior to
the explosion results in a fast runaway. The impulse of the supernova
ejecta impacting on the companion can contribute to the ejection
velocity, but for almost all the runaways considered this was a
negligible effect. The structure in this plot simply reflects the
different channels that this behaviour can follow, together with the
dependency on the mass and evolutionary state of the companion. The
sideways chevron with
$v_{\mathrm{ej}}=400\text{--}800\;\mathrm{km}\;\mathrm{s}^{-1}$ and
$B-V\sim0$ corresponds to cases in which the companion is so massive
initially that the binary is close to being equal-mass. When a more
massive star transfers mass to a lower-mass companion, the orbit
shrinks. The converse is that when a less massive star transfers mass
to a higher mass companion, the orbit grows. Thus, sustained mass
transfer causes the companion to first approach and then retreat from
the primary. The fastest runways are those in which the stars are
closest prior to the common-envelope phase and thus the tip of the
chevron represents systems in which the binary is equal-mass prior to
the common-envelope.

The binary origin of the runaway stars which escape the LMC influences their
subsequent evolution because prior to ejection more than $90\%$
experience mass transfer from the primary. The transferred mass can be
up to several $\mathrm{M}_{\odot}$ in extreme cases. Thus the runaways
in our simulation would appear as blue stragglers in comparison to
their progenitor population, i.e. would be bluer than a single star of
equivalent age and mass. If the age of a candidate runaway star is estimated using single star isochrones  and is compared to a flight time from the LMC they may be discrepant, because the rejuvenation of the star by mass
transfer prior to ejection may
have extended the lifetime of the star by a few $100\;\mathrm{Myr}$.

A finite but non-negligible time elapses between the formation of a
binary and the ejection of a runaway \citep{zapartas_delay-time_2017},
typically between $1\text{--}50\;\mathrm{Myr}$. We bin the emission
time of our runaways to the nearest $10\;\mathrm{Myr}$ because that 
is the frequency of snapshots in the N-body simulation. Once we have 
the time of ejection, we evolve each system which produces a runaway 
to the present day to ascertain the current observable properties. We 
record the stellar type, the mass, the Johnson-Cousins $UBVRIJHK$ 
magnitudes, the Sloan Digital Sky Survey $ugriz$ magnitudes and the 
\emph{Gaia} $G$, $G_{\rm BP}$, $G_{\rm RP}$ and $G_{\rm RVS}$ magnitudes. 
Because most of our binaries are B-type stars evolved on $\mathrm{Gyr}$ 
timescales, more than than 70\% of our runaways cease nuclear burning before 
the present day. If there is a supernova, we record the time that it 
occurred so we can later extract its location from the N-body model. 
We sample from a Maxwellian distribution of kick velocities for the 
neutron star and black hole remnants of Type II supernova progenitors (discussed in more detail in Section \ref{sec:exotic})
and run a second N-body integration to compute the final location of 
these compact remnants.

%
%

\begin{figure*}
	\centering
	\setlength{\fboxsep}{0pt}
	\hspace{-0.5cm}
	\begin{subfigure}{.45\linewidth}
		\begin{tikzpicture}[boximg]
		\node[anchor=south west] (img) {\includegraphics[scale=0.45,trim = 0mm 0mm 0mm 7mm, clip]{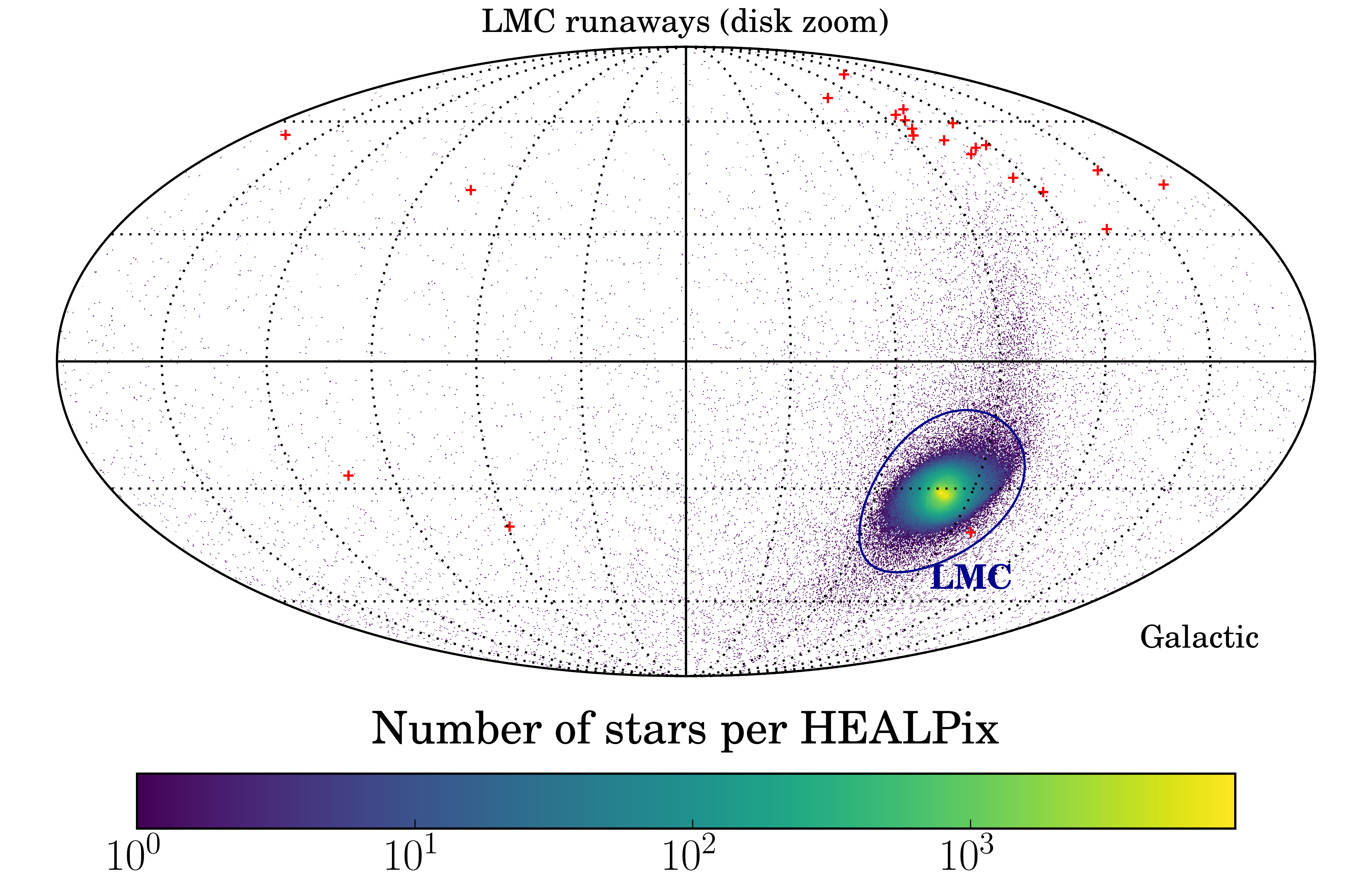}};
		\begin{scope}[x=(img.south east),y=(img.north west)]
		\node[draw,minimum height=2.4cm,minimum width=2.2cm] (B1) at (0.69,0.47) {};
		\end{scope}
		\end{tikzpicture}
	\end{subfigure}\hspace{2.1cm}
	\begin{subfigure}{.45\linewidth}
		\begin{tikzpicture}[boximg]
		\node (img1) {\fbox{\includegraphics[scale=1.20,trim = 120mm 35mm 40mm 50mm, clip]{{./allskyLMCrunawaydisk}.png}}};
		\draw (img1.south west) rectangle (img1.north east);
		\end{tikzpicture}\hfill%
	\end{subfigure}
	\hspace{-0.75cm}
	\begin{tikzpicture}[overlay,boximg]
	\draw (B1) -- (img1);
	\end{tikzpicture}
	\caption{\textbf{Left:} All-sky present day distribution of
          runaways produced by our model of the LMC. The blue circle corresponds
          to the assumed tidal radius of the LMC of
          $20\;\mathrm{kpc}$. The red crosses are the observed
          population of B-type HVSs. \textbf{Right:} Zoom-in at higher
          resolution to illustrate the structure of our LMC disk. An
          animation of the evolution of this plot through each
          snapshot of our simulation is available at
          \url{https://youtu.be/eE-1JXBP1J8}.}
	\label{fig:runaways}
\end{figure*}

\subsection{N-body MW/LMC Model}
\label{sec:nbody}

To model the runaways produced by the LMC, we use an $N$-body
simulation of the LMC and the Milky Way galaxies. The LMC is modelled
with two components (disk and dark matter halo) while the Galaxy is
modelled with three components (disk, bulge and dark matter
halo). The initial conditions are chosen such that the relative
position of the Galaxy to the LMC matches their present day value
within 2$\sigma$ (see Sec. 4 of \citealp{mackey_10_2016} for more details on the simulations). Our
simulations are evolved with the $N$-body part of \textsc{gadget-3}.
This is similar to \textsc{gadget-2}
\citep{springel_cosmological_2005} but modified in two
critical ways. First, we track the location of the centre of mass of
the LMC by using a shrinking sphere algorithm on the inner 10 kpc at
each timestep. As a consistency check, the potential
minimum of LMC particles is computed every 49 Myr and we find no
significant jumps in the LMC position. Second, the code is modified to
release massless tracer particles with a given offset in position and
velocity from the LMC. These are used to model the runaways. Before
injecting any tracers, the simulation is evolved for 1.97 Gyr to the
present and we record the LMC disk rotation curve, radial density
profile, vertical density profile, orientation, position and velocity
as a function of time. Fits to these properties, along with the extra
velocity components of runaways (described below), are used to
generate the initial conditions for the final simulation in which, as
the LMC evolves in time, tracer particles are released representing
the runaways.

The velocity vector of stars ejected from the LMC has three major
components: the orbital velocity of the LMC, the rotation of the LMC
disk and the ejection velocity of the runaway. The velocity is
dominated in most cases by the $378\;\mathrm{km}\;\mathrm{s}^{-1}$
orbital velocity of the LMC \citep{van_der_marel_third-epoch_2014}. We
initialise our runaways by sampling in cylindrical coordinates
$(R,z,\phi)$ with a weighting factor $\rho(R,z,\phi)$ which accounts
for the density of the LMC disk at each location. From the N-body
simulation, we find distributions of the tangential, radial and
vertical velocities of the stars in the LMC disk at each point in the
disk and at various times spaced at $10\;\mathrm{Myr}$. We sample
in these to determine the location and velocity of the progenitor
binary at the moment of the supernova. We then add the ejection
velocity by multiplying the ejection speed with a randomly-oriented
unit vector. The position and velocity are then converted into the
rest frame of the Galaxy.

Runaways are initialised in the simulation as massless particles every
$10\;\mathrm{Myr }$ as described in Section \ref{sec:grid} and their
orbits integrated to the present day. It is important to note that we
sample in a large number of parameters and the number of generated
runaways is relatively small. Thus, the extreme outliers of our
population are subject to small-number statistical uncertainties.

\subsection{Observables}
\label{sec:observables}

We calculate heliocentric observables for each of our runaways by
assuming that the Sun is at $R_{\odot}=8.5\;\mathrm{kpc}$ and the
Milky Way's disk rotation speed $v_{\mathrm{disk}} = 240\;
\mathrm{km}\;\mathrm{s}^{-1}$ with a solar peculiar velocity of
$(U_{\odot},V_{\odot},W_{\odot})=(11.1,12.24,7.25)\;\mathrm{km}\;\mathrm{s}^{-1}$
\mbox{\citep{schonrich_local_2010}}. We define those stars whose
present location is $20\;\mathrm{kpc}$ from the LMC to have
escaped the LMC. This is similar to the observed
$22.3\pm5.2\;\mathrm{kpc}$ tidal radius of the LMC
\citep{van_der_marel_third-epoch_2014}. A subset of the LMC escapers
will also be hypervelocity with respect to the Milky Way. We define
stars which are Galactic HVSs to be those with a Galactic rest frame
velocity greater than
\begin{align}
v_{\mathrm{esc}}(x) &=  (624.9-9.41543x+0.134835346x^2-1.292640\times 10^{-3}x^3 \nonumber \\
&+6.5435315\times 10^{-6}x^4-1.3312833\times 10^{-8}x^5) \;\mathrm{km}\;\mathrm{s}^{-1}
\end{align}
where $x=r/{1\;\mathrm{kpc}}$ and $r$ is the spherical Galactocentric radius. We take this escape velocity curve from
\cite{brown_mmt_2014} who calculated it for a three-component
potential which approximates sufficiently well our live Milky Way
Galaxy.  We then take the magnitudes from Section \ref{sec:grid},
redden them using the \cite{schlegel_maps_1998} dust map and correct
them to heliocentric apparent magnitudes. We use the present day
Cartesian coordinates of the stars to calculate the heliocentric
kinematic observables of each star including equatorial coordinates,
distance, line-of-sight velocity and proper motions.

\section{Properties of LMC Runaways}
\label{sec:prop}

The natural consequence of binary evolution in the LMC is a population
of runaway stars with extreme properties. In our simulation, tens of
thousands of stars escape the LMC with thousands surviving as
main-sequence stars at the present day. Their spatial properties and
kinematics are discussed in Section~\ref{sec:broad}.  If the LMC is as
massive as recent work suggests~\citep[e.g.][]{kallivayalil_etal_2013,penarrubia_etal_2016,jethwa_magellanic_2016}
and as is assumed in our orbital integration, then the LMC is only
marginally bound to the Galaxy and is on its first pericentre passage. A
significant fraction of the stars which are LMC escapers are also
unbound from the Galaxy, and so are HVSs. We discuss this possibility and
compare to the known population of HVSs in Section
\ref{sec:hyper}. Existing observations of a number of populations of
stars in the outskirts of the LMC lend indirect evidence to our
hypothesis, as outlined in Section~\ref{sec:misc}. The prospects for detecting an escaping LMC runaway population are discussed in Section~\ref{sec:gaia}. Lastly, a
substantial fraction of our runaway stars go supernova resulting in a
host of more exotic observables which we consider in Section
\ref{sec:exotic}. These include Type II supernovae far out in the LMC
halo, pulsars tens of kiloparsecs from the nearest site of recent star
formation and microlensing by compact remnants.

\subsection{Spatial Distribution and Kinematics}
\label{sec:broad}

The most notable feature is the extreme anisotropy of the LMC runaway
distribution on the sky, which is aligned along the orbit of the LMC
(Fig. \ref{fig:runaways}). The stars we see at a particular point
on the sky are a combination of stars which were
ejected slowly a long time ago and stars which were ejected rapidly
but more recently. The orbit of the LMC varies in heliocentric
distance and so a magnitude-limited survey will miss both low-mass
recent ejections and high-mass, high-velocity runaways that have
travelled far enough to be beyond the detection limit. We find a range
of stellar types for both LMC escapees and Milky Way HVSs (Table
\ref{tab:types}). At the present day, most of our runaways are
remnants which reflects the skew in the runaway mass distribution
introduced by the preference for high mass primaries to host high mass
companions. The lower HVS fraction of white dwarfs is because these
are the remnants of the more massive of our runaways and higher mass
stars are, to first order, ejected at lower velocities. This can be 
shown by considering the simple case of a circular binary where, if 
the mass of the primary and the separation are held constant, the orbital 
velocity of the secondary $v_2$ only exhibits a dependency on the total 
mass of the system $M$ through $v_2\propto M^{-1/2}$. Increasing the 
mass of the secondary thus decreases its orbital velocity, which in most cases is the dominant contributor to the ejection velocity. The lack of
helium white dwarfs is to be expected. Helium white dwarfs can only be
formed if the ignition of helium can be avoided, and therefore they can
only be produced from the evolution of low-mass stars over a Hubble
time or if a more massive star has its hydrogen envelope stripped by a
companion \citep[e.g.][]{althaus_evolution_1997}. Because we
specifically consider the scenario in which the companion escapes
after the explosion of the primary, the companion does not have a
chance to evolve to the giant branch and then experience mass
transfer. Conversely, if the companion remains bound to the primary
post-SN, it could then experience mass-loss as it evolves. Observed
counterparts of this channel are the well known pulsar -- helium white
dwarf binaries \citep[e.g.][]{backer_neutron_1998}. The observed single, low-mass, helium white dwarfs are instead thought to be the remnants of giant-branch donor stars whose envelope was stripped when their companion exploded as a SN Ia \citep{justham_type_2009}.

\begin{table}
	\centering
	\begin{tabular}{lrrr}
		\hline \hline Type & LMC remainers & LMC escapers & MW HVSs \\ \hline
		LM-MS & 86577 & 1227 & 77.1\%\\
		MS & 245486 & 7485 & 65.2\%\\
		HG & 1112 & 31 & 64.5\%\\
		GB & 1753 & 79 & 76.0\%\\
		CHeB & 23533 & 487 & 66.7\% \\
		EAGB & 678 & 12 & 83.3\%\\
		TPAGB & 320 & 8 & 75.0\%\\
		HeMS & 15 & 0 & ---\\
		HeHG & 2 & 0 & ---\\
		HeGB & 0 & 0 & ---\\
		HeWD & 0 & 0 & ---\\
		COWD & 510338 & 5233 & 57.0\% \\
		ONeWD & 202206 & 436 & 43.3\% \\
		NS & 146323 & 398527 & 82.9\%\\
		BH & 162646 & 445562 & 83.0\%\\ \hline 
		Total & 1380989 & 858997 & 82.6\%\\
		\hline \hline
	\end{tabular}
	\caption{Summary of stellar types at the present day by number
          of stars which either remain bound to or escape the Large
          Magellanic Cloud, and the fraction of the latter which
          are hypervelocity stars with respect to the Milky Way.
          \textbf{Key:} LM-MS - \emph{Low Mass Main Sequence}, MS -
          \emph{Main Sequence}, HG - \emph{Hertzprung Gap}, GB -
          \emph{Giant Branch}, CHeB - \emph{Core Helium Burning}, EAGB
          - \emph{Early Asymptotic Giant Branch}, TPAGB -
          \emph{Thermally Pulsating Asymptotic Giant Branch}, HeMS -
          \emph{naked Helium Main Sequence}, HeHG - \emph{naked Helium
            Hertzsprung Gap}, HeGB - \emph{naked Helium Giant Branch},
          HeWD - \emph{Helium White Dwarf}, COWD - \emph{Carbon-Oxygen
            White Dwarf}, ONeWD - \emph{Oxygen-Neon White Dwarf}, NS -
          \emph{Neutron Star}, BH - \emph{Black Hole}.}
	\label{tab:types}
\end{table}

\begin{figure*}
	\begin{tabular}{cc}
		\includegraphics[width=.48\linewidth,trim = 5mm 1mm 14mm 12mm, clip]{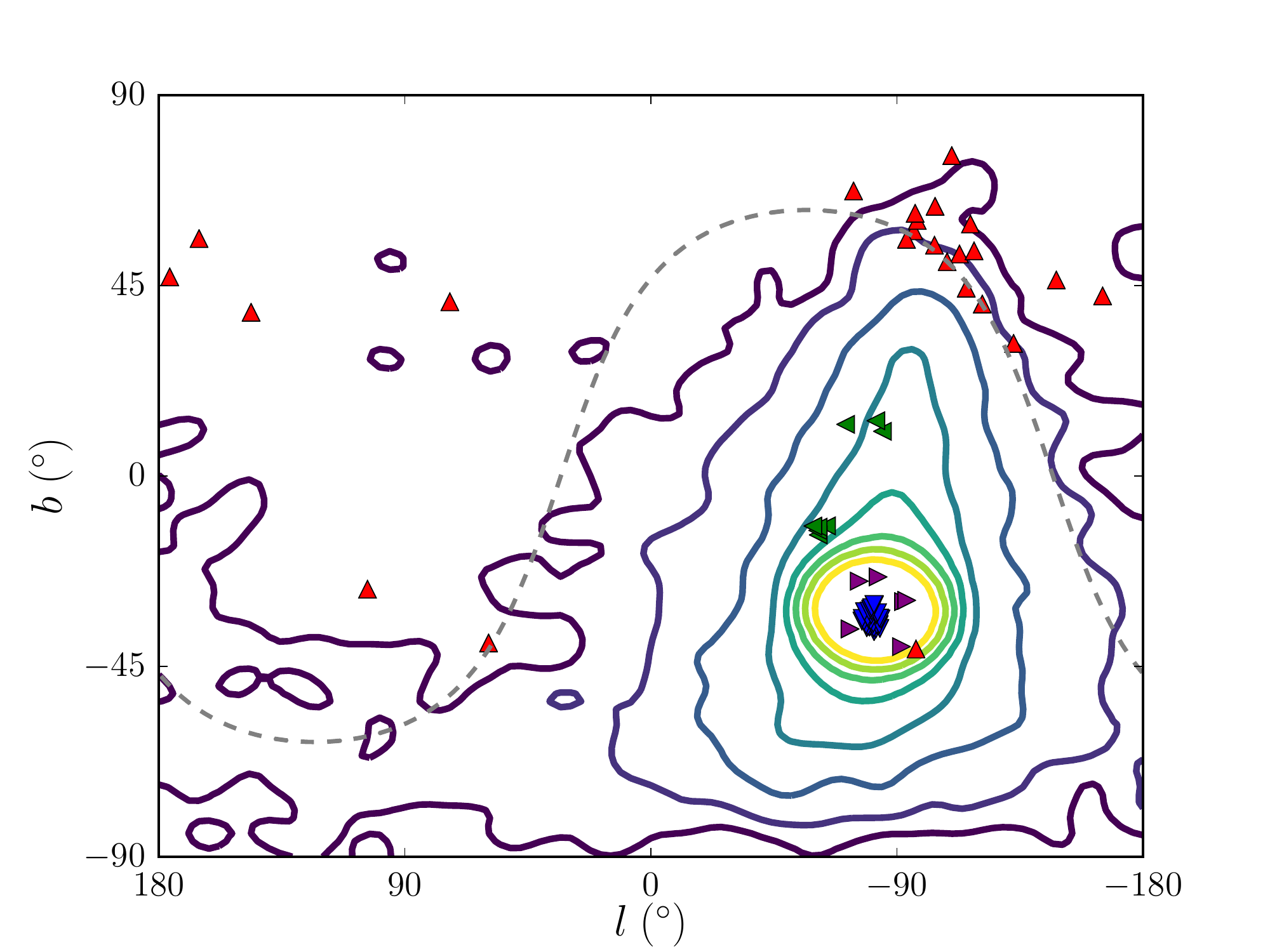} &  \hspace{-0.0cm} \includegraphics[width=.48\linewidth,trim = 5mm 1mm 15mm 12mm, clip]{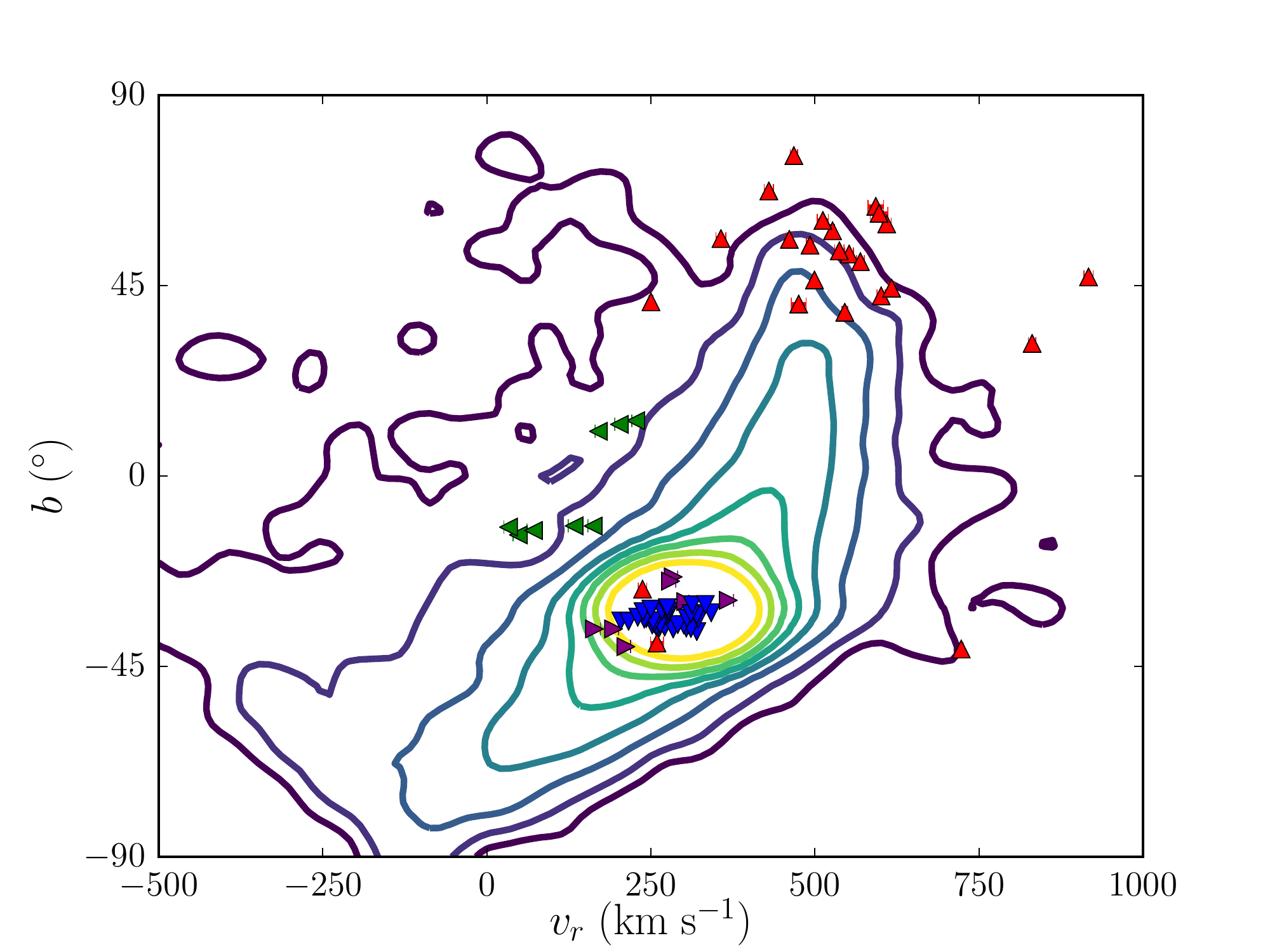} \\    (a) Latitude -- Longitude & (b) Latitude -- Radial Velocity  \vspace{0.5cm}\\
		\includegraphics[width=.48\linewidth,trim = 5mm 1mm 15mm 12mm, clip]{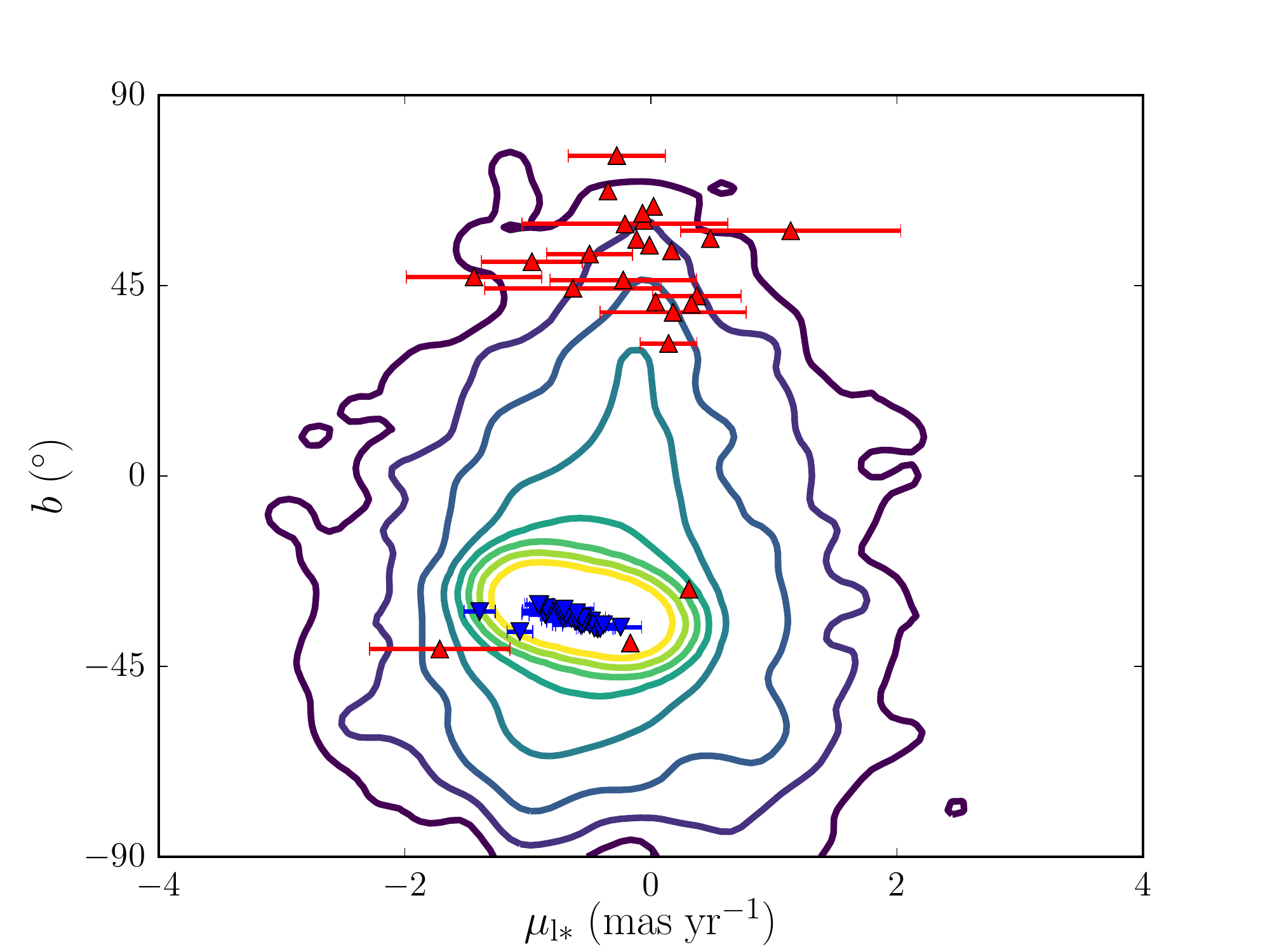} &  \hspace{-0.0cm} \includegraphics[width=.48\linewidth,trim = 5mm 1mm 15mm 12mm, clip]{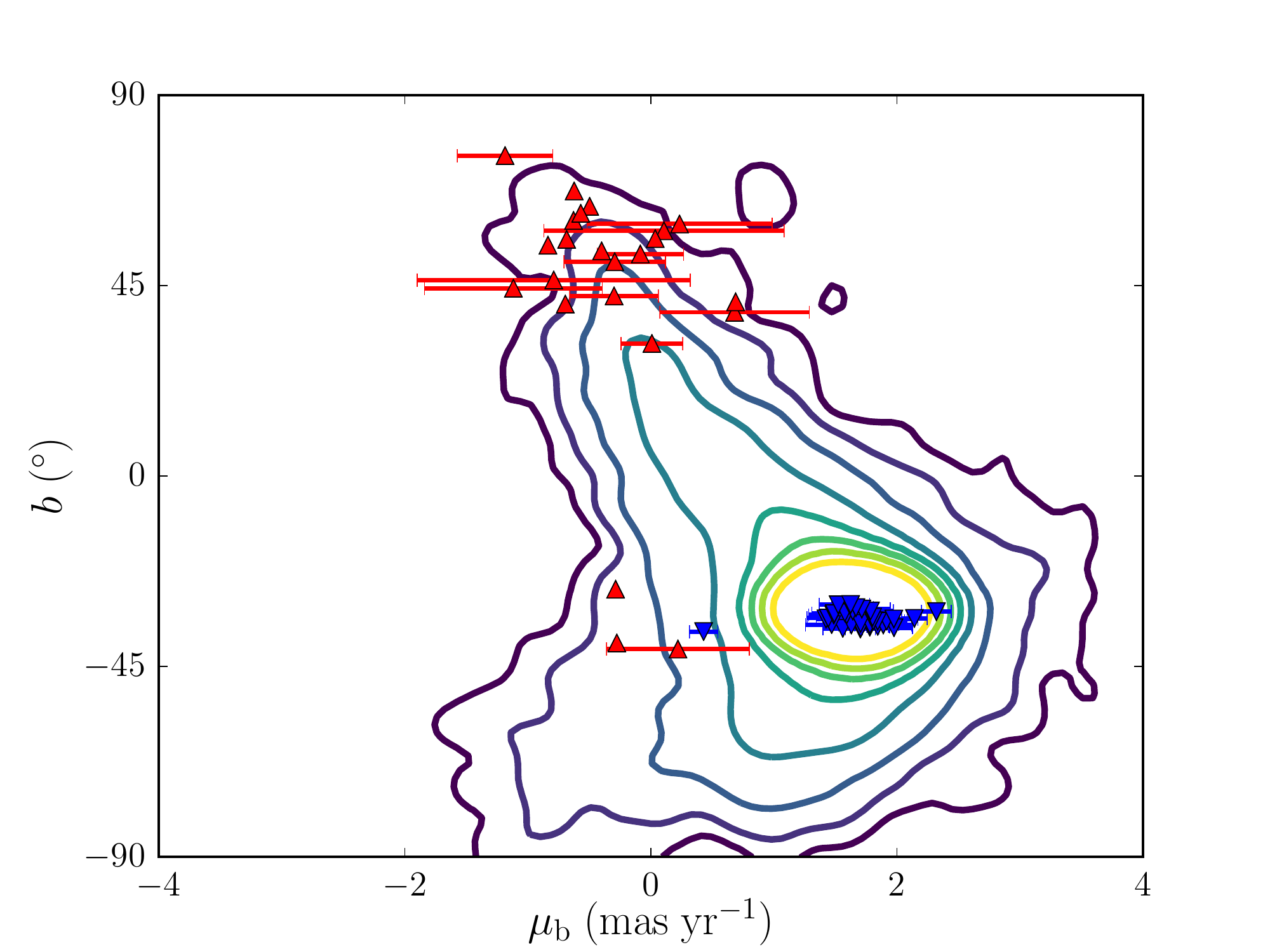} \\    (c) Latitude -- Longitudinal Proper Motion & (d) Latitude -- Latitudinal Proper Motion \vspace{0.5cm}\\
		\includegraphics[width=.48\linewidth,trim = 5mm 1mm 15mm 12mm, clip]{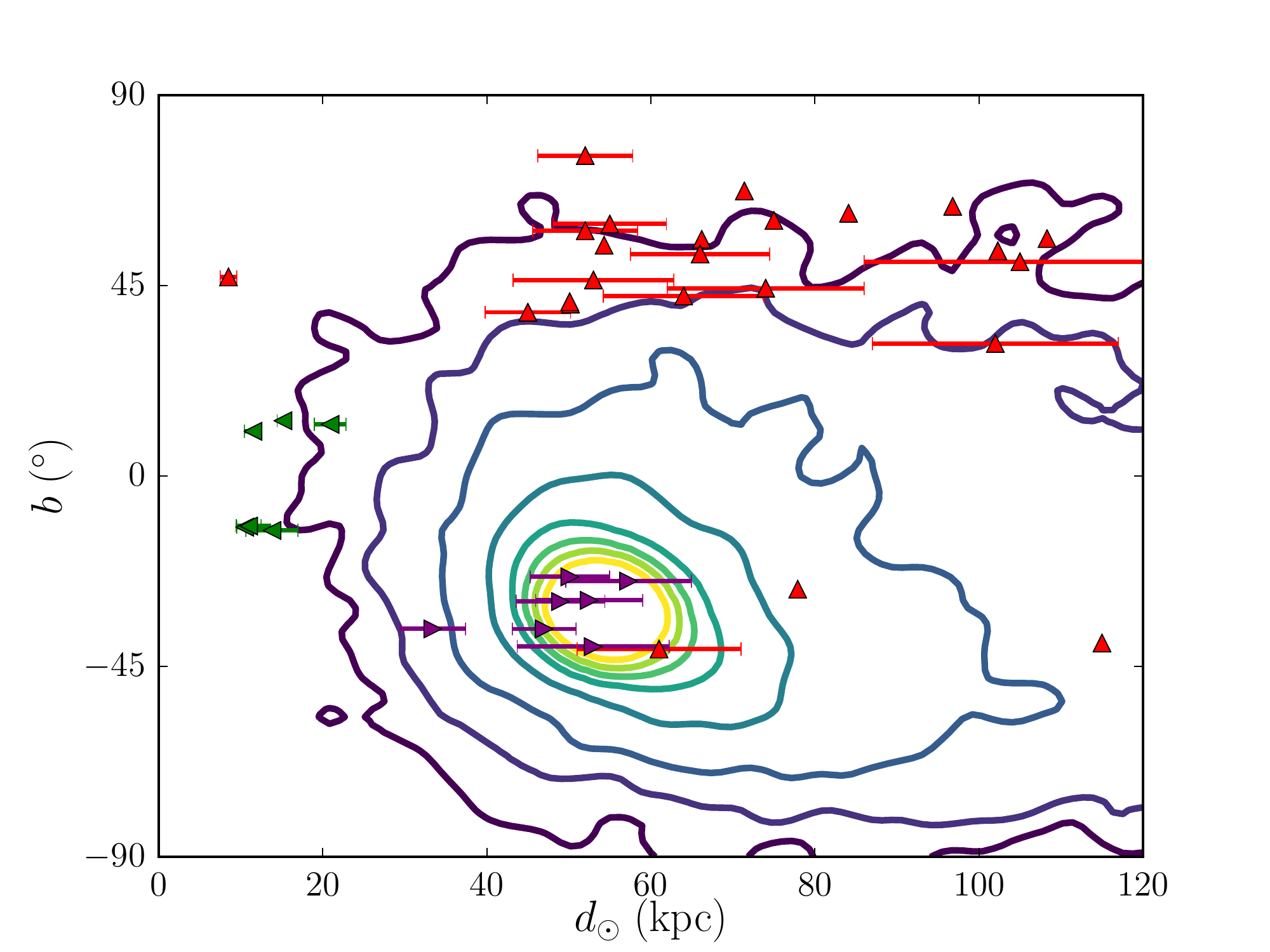} &  \hspace{-0.0cm} \begin{tabular}{c}
			\includegraphics[width=.40\linewidth,trim = 110mm 85mm 5mm 5mm, clip]{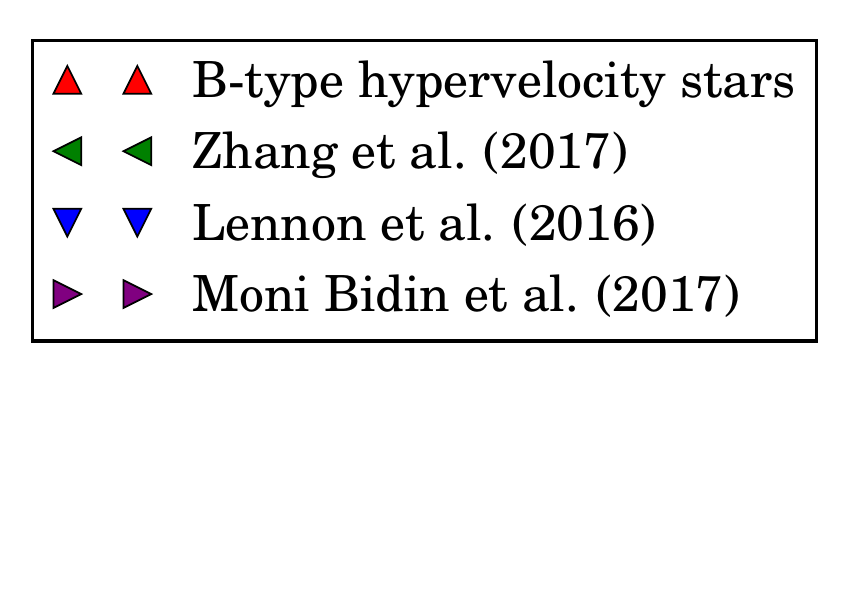} \vspace{-1.7cm} \\ \vspace{6.0cm}
			\includegraphics[width=.50\linewidth,trim = 20mm 0mm 10mm 0mm, clip]{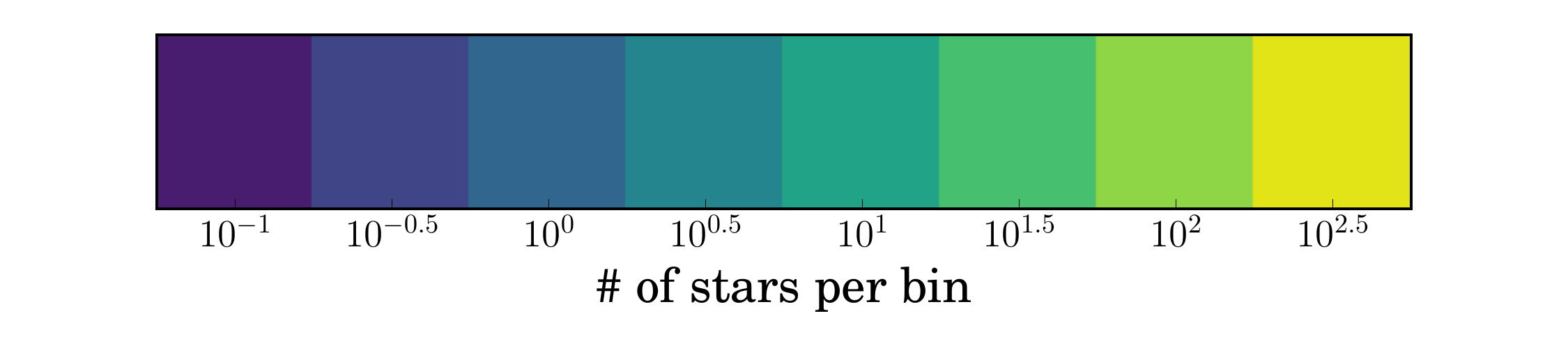}
		\end{tabular} \vspace{-5.75cm}  \\    (e) Latitude -- Distance &  \\
	\end{tabular}
	\caption{Predictions of the kinematics of our LMC runaway
          model plotted as logarithmically-spaced contours of the number of stars in each bin. The bins are defined by a $100\times100$ grid over the range of each plot. Also shown are
          observations of OB-stars near the LMC in the literature: the
          known B-type hypervelocity stars, stars which may have
          formed from the gas in the leading arm
          \citep{zhang_chemical_2017}, candidate runaways in the LMC
          \citep{lennon_gaia_2016} and young stars in the outskirts of
          the LMC \citep{bidin_young_2017}. The distances for the
          stars from \citet{zhang_chemical_2017} and
          \citet{bidin_young_2017} are calculated from distance
          moduli, proper motions were only available for the
          \citet{lennon_gaia_2016} stars and a subset of the
          hypervelocity stars, and the \citet{lennon_gaia_2016} stars
          only have distances by association with the LMC. The grey
          dashed line marks the celestial equator and SDSS photometry only covers the region above this line.}
	\label{fig:kinematics}
\end{figure*}

The orbit of the LMC is close to being polar and thus the Galactic
latitude of an LMC runaway star approximately determines its
kinematics. In Fig.~\ref{fig:kinematics}, we plot the kinematics of
the predicted LMC runaway population against Galactic latitude. We
also plot the known HVSs and several observed populations of OB-type
stars near the LMC which are discussed further in Sections
\ref{sec:hyper} and \ref{sec:misc}, respectively.

\begin{figure}
	\includegraphics[scale=0.47,trim = 6mm 0mm 10mm 10mm,
          clip]{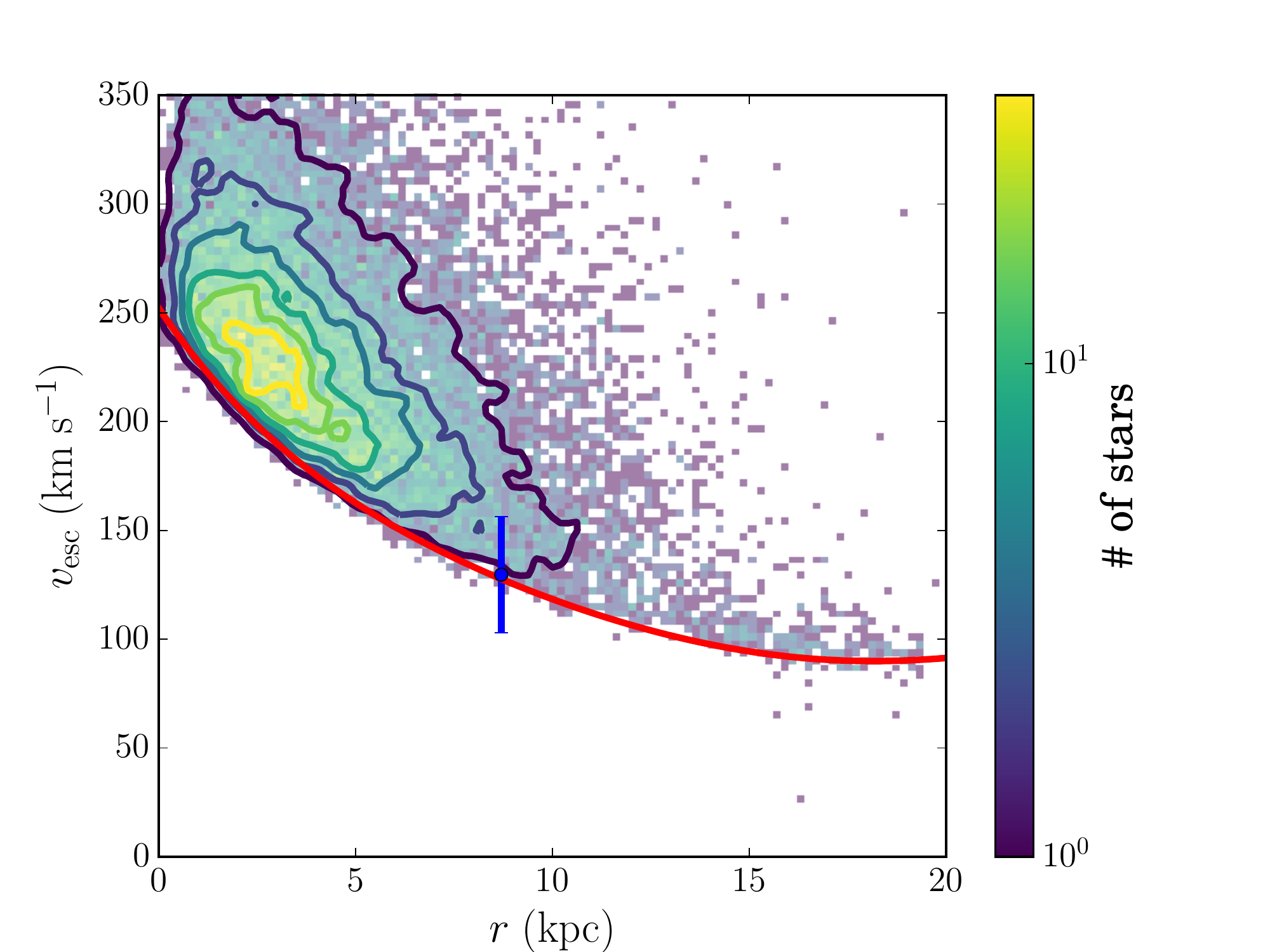}	
	\vspace{-0.2cm}
	\caption{The escape velocity curve of our modelled LMC
          potential (see Section~\ref{sec:broad}). The contours
          illustrate the distribution of our LMC escapers, the red
          line is our estimated escape velocity curve and the blue
          point is the mass constraint for the LMC
          $M(8.7\;\mathrm{kpc})=(1.7\pm0.7)\times10^{10}\;\mathrm{M}_{\odot}$
          \citep{van_der_marel_third-epoch_2014} converted to escape
          velocity with $v_{\mathrm{esc}}=\sqrt{2\mathrm{G}M/r}$, where $\mathrm{G}$ is the gravitational constant and $r$ is the spherical radius.}
	\label{fig:vesc}
\end{figure}

A convenient benefit of simulating runaway stars from a galaxy is that
it enables the calculation of the escape velocity curve, which at each 
distance from the centre of a galaxy gives the minimum speed required 
for a star at that location to be unbound. We take the
initial velocities and radii in the frame of the LMC for those stars
which we know subsequently escape to beyond $20\;\mathrm{kpc}$ from
the LMC. Because these occur sufficiently frequently at all radii
within the LMC, we estimate the escape velocity by finding
the curve that bounds these stars from below in the
$r_{\mathrm{init}}-v_{\mathrm{init}}$ plane. This is complicated by
the presence of stars which escape the LMC through the Lagrange
points, so in practice we bin the stars radially and find the first
percentile in velocity in each bin after removing outliers with
$v_{\mathrm{esc}}\leq90\;\mathrm{km}\;\mathrm{s}^{-1}$. We fit a fifth
order polynomial through these values and obtain,
\begin{align}
v_{\mathrm{esc}}(x) &=  (252.1-26.74734x +2.44534040x^2 - 0.164199176x^3 \nonumber \\
&+6.24490163\times 10^{-3}x^4-9.04817931\times 10^{-5}x^5) \;\mathrm{km}\;\mathrm{s}^{-1}, \label{eq:vesclmc}
\end{align}
where $x=r/1\;\mathrm{kpc}$ and $r$ is the spherical radius from the LMC centre, which we plot in
Fig.~\ref{fig:vesc}. Note that because we have a lower initial density
of stars at large radii, the escape velocity curve is less accurate at
these distances and we would not advocate using it outside
$15\;\mathrm{kpc}$. Eqn.~\ref{eq:vesclmc} is the escape velocity curve
of the LMC in isolation. The LMC has been truncated by the Milky Way
at the tidal radius by the present day and thus the escape velocity
currently is lower than over the previous $1.97\;\mathrm{Gyr}$.

\subsection{Hypervelocity Stars (HVSs)}
\label{sec:hyper}

If all HVSs originate in the Galactic Centre, we expect them to be
isotropically distributed on the sky. However,
\citet{brown_anisotropic_2009} found that eight of the 14 HVSs in the
\citet{brown_hypervelocity_2007,brown_mmt_2009} targeted surveys are
in the constellations of Leo and Sextans, despite the surveys covering
one fifth of the sky. This anisotropy is not simply a selection
effect, since \citet{brown_hypervelocity_2007} is 100\% complete for
stars with $17<g_0'<19.5$ over the 7300 deg$^2$ covered by the Sloan
Digital Sky Survey Data Release 6 and \citet{brown_mmt_2009} is 59\%
complete for stars with $19.5<g_0'<20.5$ over the same
region. \citet{brown_anisotropic_2009} attempted to verify the
significance of the anisotropy by showing that the HVSs are clustered
compared to the stars in the surveys in both Galactic latitude and
longitude at $3\sigma$ significance, in angular separations at
$5\sigma$ significance and in two-point angular correlation at
$\sim3.5\sigma$ significance. \citet{brown_hypervelocity_2015} states
that there is currently ``no good explanation for the anisotropic
distribution of unbound late B-type
stars". \citet{boubert_dipole_2016} suggested that this anisotropy
could be explained by Hills ejection of stars by a currently
undetected SMBH at the centre of the LMC.

An LMC origin had previously been explored for the one HVS in the
southern hemisphere, HE 0437-5439, which was discovered by
\cite{edelmann_he_2005}. The flight time is longer than the
main-sequence lifetime of the star and hence either it is a blue
straggler, and was ejected as a binary from the Galactic Centre, or it
has its origin in the LMC. The mechanism that ejected HE 0437-5439
from the LMC has been suggested to be either interactions with a black
hole more massive than
$10^3\;\mathrm{M}_{\odot}$\citep{gualandris_hypervelocity_2007} or
dynamical ejection from a cluster \citep{przybilla_lmc_2008}.

In this work, we consider the population of HVSs produced by the binary
supernova runaway mechanism operating in the LMC, which Table
\ref{tab:types} demonstrates is substantial. However, we find that
our model LMC runaway HVSs which make it into the footprint of SDSS are
inconsistent with the observed HVSs, being in the mass range
$1.6\;\mathrm{M}_{\odot}<M<3.0\;\mathrm{M}_{\odot}$ rather than the
$M>3.0\;\mathrm{M}_{\odot}$ of the B-type HVSs. The reason for this is
clear from Fig.~\ref{fig:bvvej}. Those stars that make it into the
footprint of SDSS have $v_{\mathrm{ej}}\gtrsim
200\;\mathrm{km}\;\mathrm{s}^{-1}$ and there is a distinctly low
probability density of runaways at these speeds with $B-V<0$. There are
three possibilities that either dismiss or resolve this discrepancy:
\begin{enumerate}
	\item The observed B-type HVSs do originate in the Milky Way
          galaxy from one of the processes discussed above and the
          anisotropy indicates a symmetry-breaking in these
          processes. One example would be if the binary stars which
          interact with Sgr A$^{*}$ are scattered from a disk in the
          Galactic nucleus rather than coming from a
          spherically-symmetric population.
	\item The observed B-type HVSs originate in the LMC, but are
          ejected by a process which has a higher typical ejection
          velocity than runaways -- either the Hills mechanism or
          dynamical ejection from a cluster.
	\item Our prescription for the common-envelope evolution of
          binary stars is inaccurate. We follow
          \citet{hurley_evolution_2002} and set
          $\alpha_{\mathrm{CE}}=1.0$, where $\alpha_{\mathrm{CE}}$ is
          the efficiency with which the orbital energy of the binary
          can be used to remove the common envelope, but this
          parameter is not well-constrained observationally. If we
          instead set $\alpha_{\mathrm{CE}}=0.1$, we find a
          high-velocity distribution where the HVSs would be
          predominantly of A and B type. There is additional
          uncertainty in the fraction $\lambda_{\mathrm{CE}}$ of the
          binding energy of the envelope which is required to eject
          the envelope. We use a fit to tabulated numerical results
          which is implemented in {\sc binary\_c}
          \citep{dewi_energy_2000,tauris_binding_2001}. However, the
          tabulated $\lambda_{\mathrm{CE}}$ were calculated at solar
          metallicity. The parameters $\alpha_{\mathrm{CE}}$ and
          $\lambda_{\mathrm{CE}}$ appear together in the
          $\alpha$-prescription and so it is the combination
          $\alpha_{\mathrm{CE}}\lambda_{\mathrm{CE}}$ which sets the
          post-common-envelope separation. An error in either
          parameter could explain the possible discrepancy between the
          observed HVSs and our LMC runaway model.
\end{enumerate}

We can evaluate whether the observed HVSs originate in the LMC,
whilst being agnostic about the mechanism, by considering the runaway
stars in our model to be tracer particles of the kinematic
distribution of stars ejected from the LMC. When discussing the known HVSs we specifically refer to the candidates discovered by the HVS surveys \citep{brown_discovery_2005,brown_successful_2006,brown_hypervelocity_2007-1,brown_hypervelocity_2007,brown_mmt_2009,brown_mmt_2012,brown_mmt_2014} in addition to HE 0437-5439 \citep{edelmann_he_2005} and US 708 \citep{hirsch_us_2005}, with recent updated proper motions  from the Hubble Space Telescope \citep{brown_proper_2015}. Fig.~\ref{fig:kinematics}
demonstrates that the 6D kinematics of the known HVSs are consistent
with the expectations for an LMC origin. The agreement in proper
motions and distance is not surprising. The known HVSs were found in
observation campaigns
\citep{brown_successful_2006,brown_hypervelocity_2006,brown_mmt_2009}
that selected for distant B-type stars in the footprint of SDSS, and
thus most have $\delta>0^{\circ}$ and are at distances
$50<d<120\;\mathrm{kpc}$. At these distances, the proper motion
projects to nearly zero independent of whether the star originates in
the Galaxy or LMC. It is surprising, however, that an LMC origin can
reproduce the clustering in the $b\text{--}l$ and $b\text{--}v_r$
plots, neither of which can be explained by a spherically-symmetric
ejection from the Galactic Centre by the Hills mechanism. In Fig.~\ref{fig:kinematics} (a), we include a dashed line equivalent to $\delta=0^{\circ}$ which corresponds to the lower edge of the region of the sky which has been thoroughly searched for HVSs. The current searches for HVSs using SDSS are in the wrong part of the sky for the majority of an LMC escaping distribution. The other populations
of OB stars shown in Fig.~\ref{fig:kinematics} are from comparatively
shallow surveys down to magnitudes around $V=16\;\mathrm{mag}$, while
the known HVSs have SDSS magnitudes in the range $17.5<g_0<21.0$. If
the observed HVS population does originate in the LMC, then the final
\emph{Gaia} catalogue, complete down to $G\approx20.7\;\mathrm{mag}$, could
contain hundreds or even thousands of stars which have escaped the
LMC, the majority of which would be HVSs.

\subsection{Observations of Outer LMC Populations}
\label{sec:misc}

Recently, \cite{zhang_chemical_2017} reported high-resolution spectra
of eight previously claimed candidates
\citep{casetti-dinescu_constraints_2012,casetti-dinescu_recent_2014}
for OB-type stars which have formed from the gas in the Leading Arm of
the Magellanic System. They found that for five of these stars their
chemistry was consistent with an LMC origin and that their kinematics
appeared to rule out membership of the Milky Way
disk. \cite{zhang_chemical_2017} concluded that these stars therefore
must have formed from the gas in the leading arm. One property of
these stars is however quite puzzling: none display a clear signal of
radial velocity variation from a binary
companion. \cite{zhang_chemical_2017} factor in the detection
efficiency of their observations and calculate that the probability of
their null detection is 14\% (8.7\%) if the underlying binary
fraction is 50\% (60\%). While this is not statistically significant
evidence for an unusually low binary fraction, the null detection of
companions is entirely consistent with our prediction of B-type
runaway stars from the LMC. \citet{casetti-dinescu_recent_2014}
rejected a Galactic runaway origin for these five B-type stars arguing
that their radial velocity dispersion of
$33\;\mathrm{km}\;\mathrm{s}^{-1}$ is too low compared to the $\sim
130\mathrm{km}\;\mathrm{s}^{-1}$ \citep{bromley_runaway_2009} expected
for Milky Way runaways, and that an ejection mechanism would need to
be ``directionally coherent, which is highly unlikely". However,
\citet{casetti-dinescu_recent_2014} do not consider a runaway origin
from the LMC which naturally explains the low velocity
dispersion. There is one O6V star, labelled by \citet{zhang_chemical_2017} as CD14-A08, which
\citet{casetti-dinescu_recent_2014} do consider as originating in the
LMC, but they argue it must have formed in-situ from the gas of the Leading Arm
since its lifetime is too short ($1\text{--}2\;\mathrm{Myr}$) for it
to have travelled from the LMC at any less than about $10^{4}\;\mathrm{km}\;\mathrm{s}^{-1}$. However, \citet{zhang_chemical_2017}
argue that CD14-A08 is more likely to be a helium-deficient sdO star with 
$\log N_{\mathrm{He}}/N_{\mathrm{H}}=-1.69\pm0.24$. \citet{martin_kinematics_2017} 
discuss the likely production mechanism of subdwarf stars as a function of their 
helium abundance. Helium-deficient subdwarfs are thought to be produced by close 
interactions in a binary and have ages between $0.2$ and $10\;\mathrm{Gyr}$, allowing 
CD14-A08 to have originated anywhere in the MW or the LMC. 
\citet{martin_kinematics_2017} mention the possibility that intermediate-helium 
sdO/sdB stars are the polluted, runaway companions of SN Ia progenitors, which has previously been used to explain the helium-rich HVS US708 
\citep{justham_type_2009,geier_progenitor_2013}. This suggests an intriguing 
alternative origin for CD14-A08 as a runaway from a Type Ia SNe in the LMC, which may 
be required if more precise data constrain the helium abundance to be in the range 
$5\%<n_{\mathrm{He}}<80\%$ considered by \citet{martin_kinematics_2017} to be 
intermediate-helium. In Fig.~\ref{fig:kinematics}, we show the kinematics of
the \citet{zhang_chemical_2017} sample against the LMC runaway
predictions. These stars are consistent with an LMC runaway
origin. Their position near the edge of the LMC runaway distribution
in radial velocity and distance is a natural consequence of the
shallowness of the survey, which only probes the nearest edge of the
distribution in regions where we would predict relatively low radial
velocities.

\cite{lennon_gaia_2016} combined the precise proper motions of the
Tycho Gaia Astrometric Solution (TGAS) with prior radial velocity
surveys to search for runaway stars amongst the 31 brightest stars in
the LMC. They found that only two of these 31 candidates are outliers
in velocity, while the remaining stars are consistent with a rotating
disk. In fact, the majority of our runaways would be classed as
walkaways, with 65\% of runaways having ejection velocities less than
$10\;\mathrm{km}\;\mathrm{s}^{-1}$, and hence indistinguishable from
the disk population. There is also the statistical argument that most
massive stars are in binaries, so most of these stars are either
runaways or have a companion. Of the two outliers, Sk-67 2 is
suggested as a candidate hypervelocity star based on a peculiar
velocity of $359\;\mathrm{km}\;\mathrm{s}^{-1}$ and R 71 could be the
evolved product of a slow runaway binary. Note that R 71 is a Luminous
Blue Variable (LBV). It was hypothesised by \cite{smith_luminous_2015}
that the higher spatial dispersion of LBVs versus O-type and
Wolf-Rayet stars in the LMC indicates either that LBVs are merged stars or they are runaway stars that were rejuvenated by mass transfer before being ejected. This contradicts
the standard view of LBVs as a necessary transition state of massive
stars between core hydrogen burning and the Wolf-Rayet phase. We seek
analogues of the runaway candidates of \cite{lennon_gaia_2016} in our
simulation, assuming they lie at a distance of $50.1\pm3.0
\mathrm{kpc}$, and find that most are consistent with a runaway origin
(Fig.~\ref{fig:kinematics}). We are hindered because we
compare the brightest stars between observations and our model LMC runaway population. Small number statistics dominate and it
is difficult to quantify whether any particular star is inconsistent
with our model. The hypervelocity candidate Sk-67 2 is the clear
outlier from the other candidates of \citet{lennon_gaia_2016} in
Fig.~\ref{fig:kinematics} (d) where we plot
$b\text{--}\mu_{\mathrm{b}}$. It is possible that the Hills
mechanism or dynamical ejection is required to explain this star. The
other outlier in Fig.~\ref{fig:kinematics} (c) is Sk-71 42 which
\citet{lennon_gaia_2016} note as having a large
astrometric\_excess\_noise parameter in TGAS and stated that further
data would be necessary before they could speculate on the nature of
the star.

It is interesting to note the similarities between Sk-67~2 and a previous discovery by \citet{evans_runaway_2015} of a $12\text{--}15\;\mathrm{M}_{\odot}$ runaway red supergiant J004330.06+405258.4 at a projected distance of $4.6\;\mathrm{kpc}$ from the plane of M31's disk. \citet{evans_runaway_2015} mention that J004330.06+405258.4 may be a high-mass analogue of the MW HVSs since it is likely unbound from M31. Both stars are supergiants and both are discrepant with their host galaxies' kinematics by $\sim300\;\mathrm{km}\;\mathrm{s}^{-1}$. \citet{evans_runaway_2015} mention four previous discoveries of yellow and red supergiants in the LMC, SMC and M33 which have peculiar velocities around $150\;\mathrm{km}\;\mathrm{s}^{-1}$. These massive runaways are difficult to reproduce in our current model, however a modification of the common-envelope prescription to produce more early-type stars would likely resolve this problem (Sec. \ref{sec:hyper}). These stars are some of the brightest stars visible in the Local Group and so are obvious candidates for spectroscopic follow-up when they are found far from central star formation regions. It is possible that these stars are only the first tracers of a high-velocity runaway population which exists throughout the Local Group. 

\citet{bidin_young_2017} searched for star formation on the periphery
of the LMC disk between $6^{\circ}$ and $~30^{\circ}$ from the
centre. They found six recently formed stars well away from the central
star formation in the LMC, with $V<16$, separation
$7^{\circ}$--$13^{\circ}$ and ages between $10$ and
$50\;\mathrm{Myr}$. They argued that if their tangential velocity is
only as discrepant from the LMC disk tangential velocity as their
radial velocity component, these stars cannot have travelled to their
current location within their lifetimes. However, in our simulation,
analogues of these stars do exist with similar ages because the
assumption of equally discrepant velocity components does not
hold. The existence of a ring-like structure is a natural consequence
of sampling a small number of stars from a population which rapidly
decreases in number with radius and is truncated at $6^{\circ}$ from
the LMC.

\subsection{Prospects with Gaia}
\label{sec:gaia}

\begin{figure}
	\begin{tabular}{c}
		\includegraphics[width=.96\linewidth,trim = 5mm 1mm 14mm 12mm, clip]{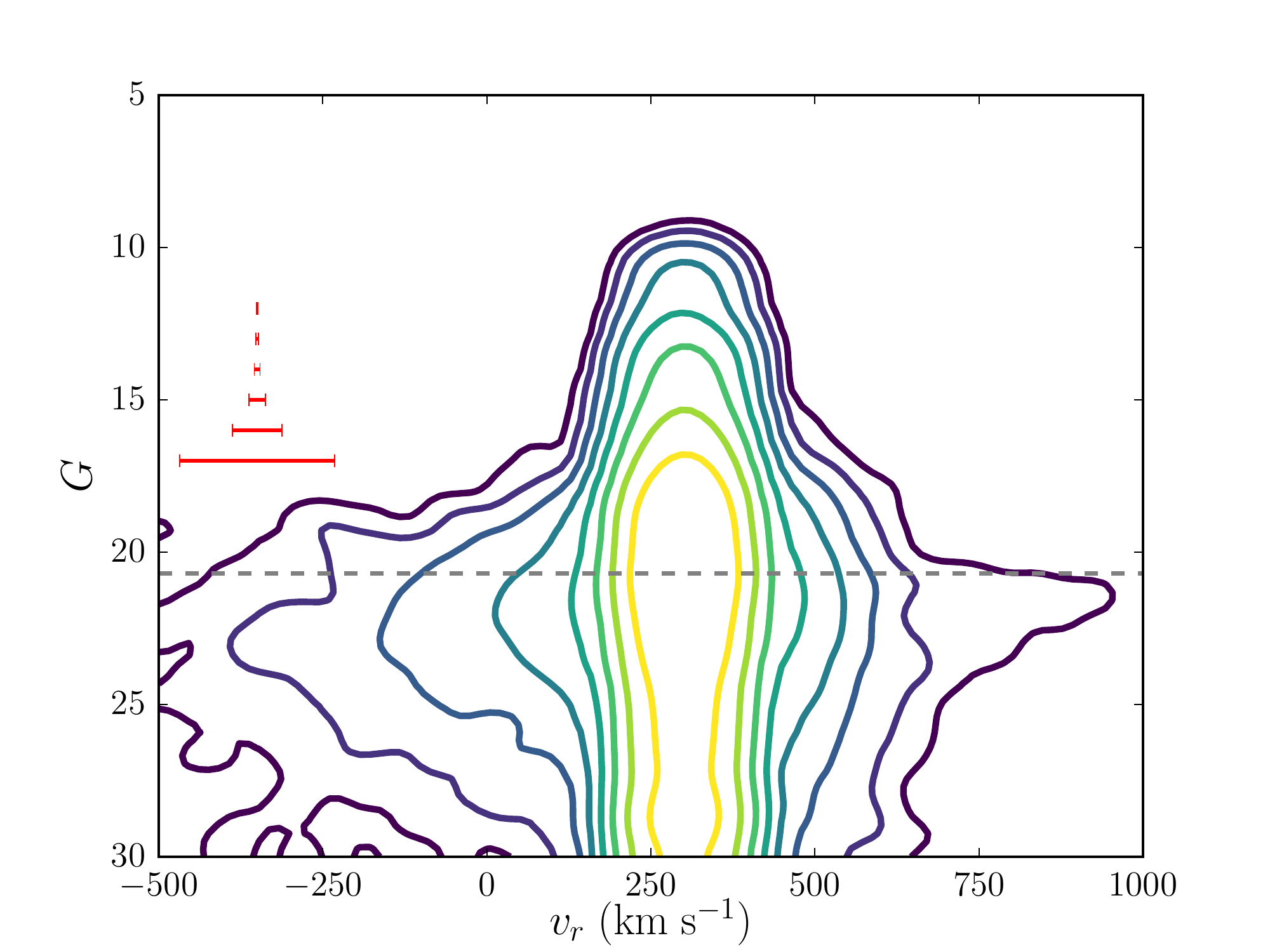} \\    (a) Apparent Magnitude -- Radial Velocity \vspace{0.5cm}\\
		\includegraphics[width=.96\linewidth,trim = 5mm 1mm 15mm 12mm, clip]{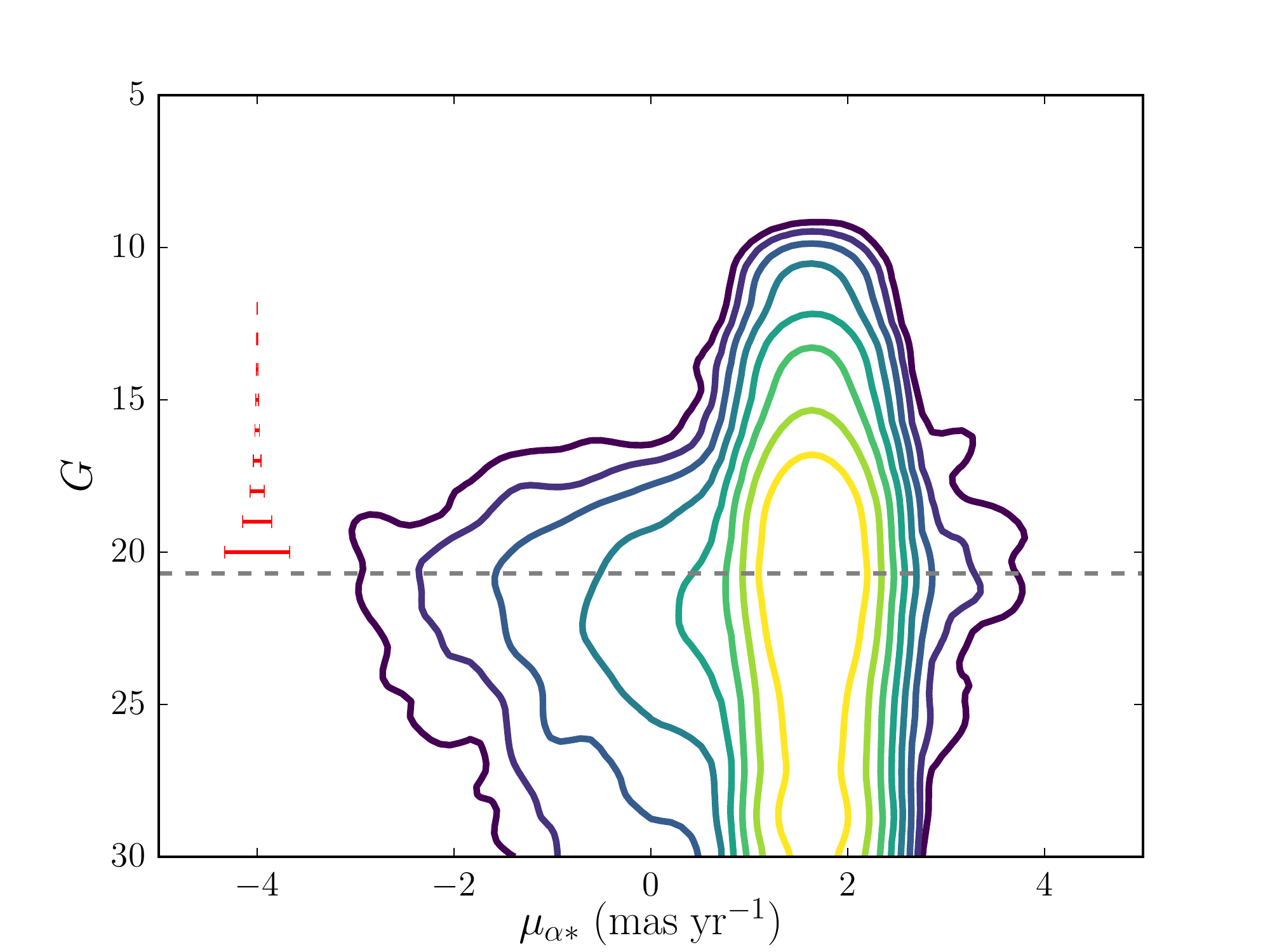} \\    (b) Apparent Magnitude -- Longitudinal Proper Motion  \vspace{0.5cm}\\
		\includegraphics[width=.96\linewidth,trim = 5mm 1mm 15mm 12mm, clip]{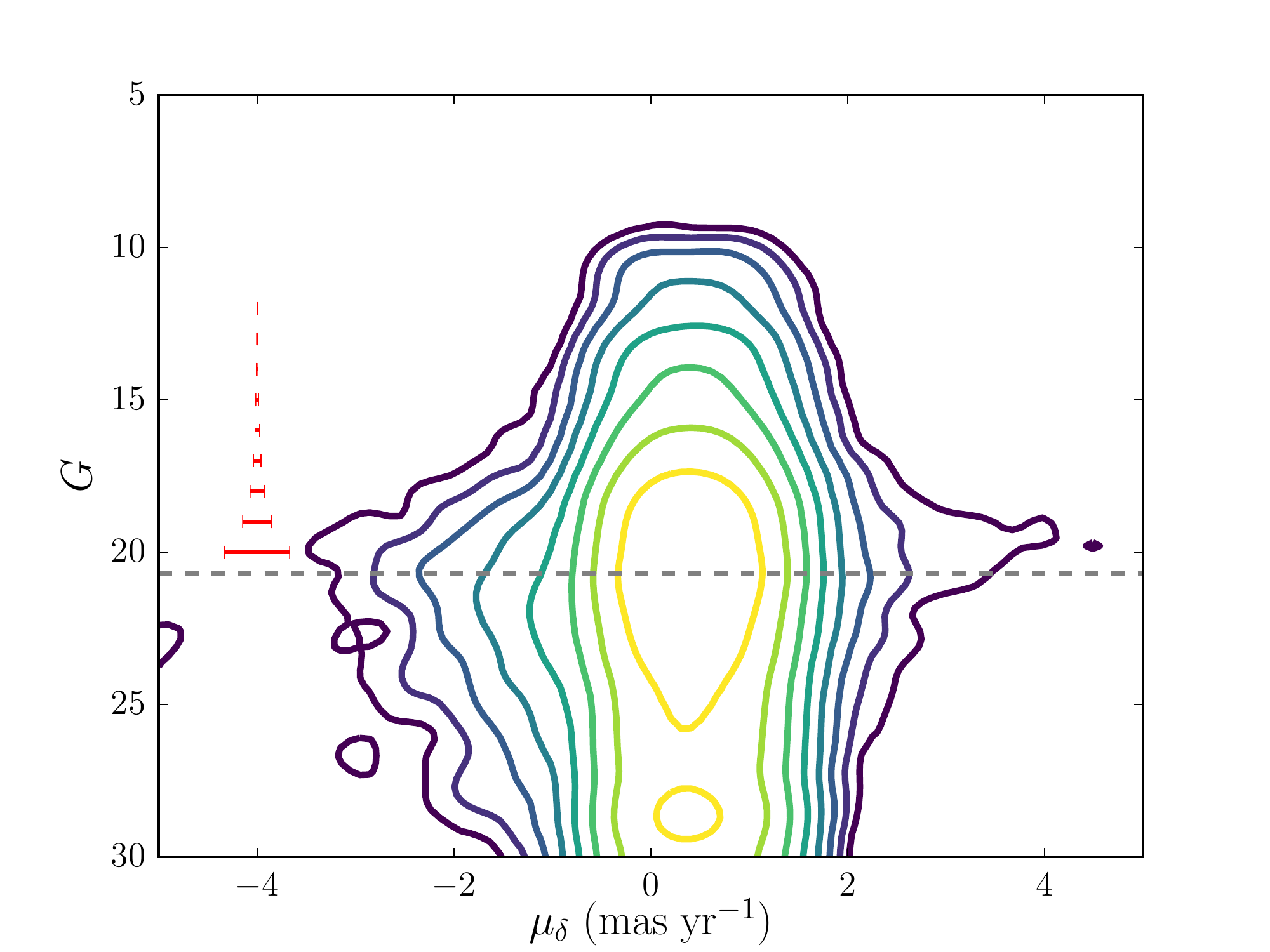} \\    (c) Apparent Magnitude -- Latitudinal Proper Motion \\
	\end{tabular}
	\caption{Predicted properties of LMC runaways which
          would be observed by \emph{Gaia} plotted as
          logarithmically-spaced contours of the number of stars in
          each bin (see Fig. \ref{fig:kinematics} for the
          colourbar). The kinematics are heliocentric and $G$ is the
          unreddened apparent magnitude. The grey dashed line
          indicates the $G\approx20.7$ completeness limit for
          \emph{Gaia} and the red error bars represent the
          $\pm1\sigma$ predicted end-of-mission radial velocity and
          proper motion errors as a function of $G$ (described in detail in
          Sec. \ref{sec:gaia}).}
	\label{fig:gaia}
\end{figure}

The \emph{Gaia} satellite is predicted to be complete down to
$G\approx20.7$, hence will be the first survey covering the Southern
hemisphere which is sensitive to the population of runaway stars which
may have escaped the LMC. We compare the predicted observable
properties of the LMC runaways to the expected $\pm1\sigma$
end-of-mission radial velocity and proper motions errors for
\emph{Gaia} in Fig.~\ref{fig:gaia}. The proper motion errors are the
predicted sky-average errors for an unreddened G2V
star\footnote{\url{https://www.cosmos.esa.int/web/gaia/sp-table1}}. The
radial velocity errors are calculated for an unreddened G0V star using
a standard performance
model\footnote{\url{https://www.cosmos.esa.int/web/gaia/science-performance}}
which is valid down to $G_{\mathrm{RVS}}\sim16$, where we used the
colour-colour relations calculated by \citet{jordi_gaia_2010} to
convert $G$ to Johnson $V$ and $G_{\mathrm{RVS}}$. The mean mass of
the LMC escapers is $1.35\;\mathrm{M}_{\odot}$ which justifies the
choice of G0V/G2V to illustrate the errors, however there are a range
of LMC escaper masses. More (less) massive stars will have larger
(smaller) errors. The radial velocities measured by \emph{Gaia} are unlikely
to have the necessary precision to detect the population of escaping
LMC runaways (Fig. \ref{fig:gaia} (a)), with the possible
exception of the bright $G=15\text{--}16$ and fast $v_r\approx
500\;\mathrm{km}\;\mathrm{s}^{-1}$ stars. Figures \ref{fig:gaia} (b)
and (c) show that the $\mu$as astrometric precision of \emph{Gaia} should
result in the detection of high velocity runaways purely by their
proper motion. The uncertainties on the parallax measurements by \emph{Gaia}
rule out the possibility of a significant detection of parallax in
LMC runaway stars. Distances would need to be obtained photometrically
to validate any candidates. The LMC escapers will also be distinct
from the LMC in their position on the sky and thus we conclude that
\emph{Gaia} will observe such a population if it exists. A change in the
common envelope prescription to produce more early-type stars 
(Sec. \ref{sec:hyper}) would not change this conclusion because the
small increase in the astrometric uncertainties at fixed $G$ is
more than cancelled by the shift of the distribution to brighter $G$
magnitudes.

\subsection{Exotica: Runaway Supernovae, Pulsars and Microlensing}
\label{sec:exotic}

\subsubsection{Runaway Supernovae and Pulsars}

\begin{figure*}
	\includegraphics[scale=0.85,trim = 8mm 0mm 8mm 7mm, clip]{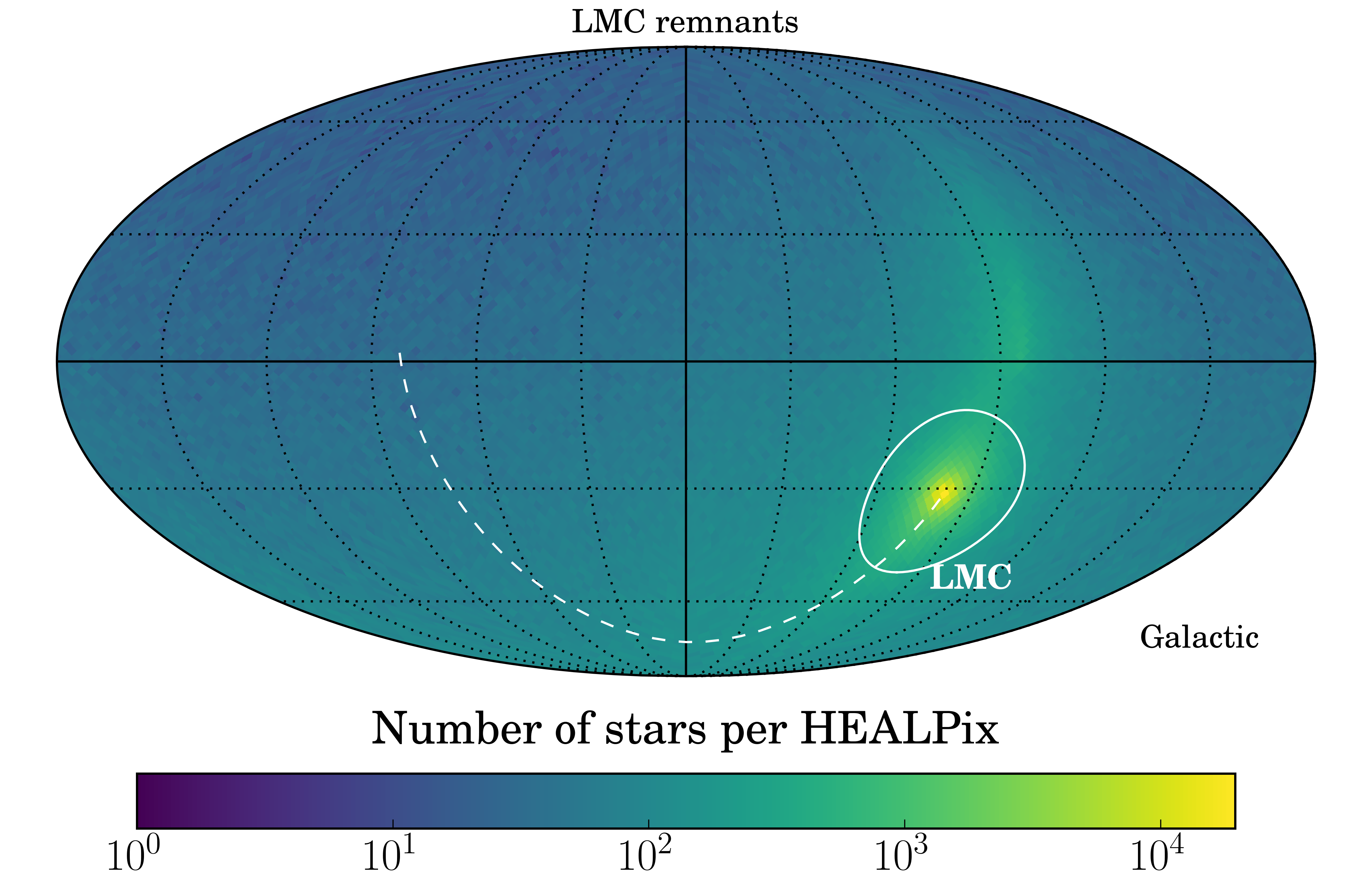}
	
	\vspace{-0.1cm}
	
	\caption{All sky distribution of remnants produced by runaway
          supernovae in our models (Sec. \ref{sec:exotic}). The white
          solid line indicates the $20\;\mathrm{kpc}$ tidal radius of
          the LMC and the white dashed line is the orbit of the LMC
          over the last $1.97\;\mathrm{Gyr}$ in the frame where the
          Sun is fixed at $(x,y,z)=(-R_{\odot},0,0)$.}
	\label{fig:remnants}
\end{figure*}

In our model, a substantial fraction of runaways (51.0\%) have experienced a core-collapse supernova before the present day, at a rate of
$5.9\times10^{-4}\;\mathrm{yr}^{-1}$, leaving behind a compact neutron
star or black hole remnant. The compact remnants experience a kick
which we prescribe to be Maxwellian-distributed with a dispersion of
$190\;\mathrm{km}\;\mathrm{s}^{-1}$ \citep{hansen_pulsar_1997}.
However, the fact that pulsars exist in globular clusters suggests
that a fraction of neutron stars could receive almost no kick at birth
\citep{podsiadlowski_neutron-star_2005}. Several authors have found
that a bimodal Gaussian is required to describe the observed pulsar
velocity distribution
\citep{fryer_population_1998,cordes_neutron_1998}, but these studies
differ on the required properties of such a distribution. Given that
the runaway velocity distribution is itself uncertain, we feel
justified in preferring the simplicity of a unimodal distribution in
this study. The SN kick, in most cases, dominates the velocity of the
remnant. The majority of these remnants subsequently escape the LMC
and most of those are unbound from the Galaxy
(Fig.~\ref{fig:remnants}). Despite the high kick dispersion, the
distribution on the sky preserves the signal of their LMC origin and
thus, if they are observable, their origin is unambiguous. There are
few accessible observables associated with single, compact remnants at
tens of kiloparsecs. However, for the first few tens of millions of
years, neutron stars manifest themselves as pulsars.

The Australia Telescope National Facility Pulsar Catalogue
\citep[][available at
  \url{http://www.atnf.csiro.au/research/pulsar/psrcat}]{manchester_australia_2005}
reveals there are ~29 pulsars currently associated with the LMC or
SMC. We cannot accurately estimate the distance to these pulsars
except through their plausible association with the Magellanic
Clouds. For pulsars too far away for parallax measurements, the
primary distance estimate is found by relating the dispersion measure
to the integrated electron column density along the line of
sight. This method is only reliable out to distances of $\sim
20\;\mathrm{kpc}$. For example, the most recent electron density maps
made by \cite{yao_new_2017} return a maximum distance of
$25\;\mathrm{kpc}$ to any pulsar with an anomalously high dispersion
measure. There are 75 pulsars in our simulation closer than this upper
limit. However, the completeness of the existing pulsar surveys is
patchy at these distances, and all but one of the pulsars estimated to
lie beyond $20\;\mathrm{kpc}$ are in the direction of the well-studied
Galactic bulge. The wide field of view and high sensitivity of the
Square Kilometre Array will enable the discovery of 20,000 new pulsars
\citep{smits_pulsar_2009}. This is an order of magnitude increase in
sample size and will provide a test of our model. The possibility that
hundreds of thousands of neutron stars have been ejected from the LMC
and are now populating the local IGM was mentioned by
\cite{ridley_new_2010} in the context of single star evolution.


\subsubsection{Microlensing}
\label{sec:microlensing}

Photometric microlensing towards the LMC by an intervening population
of dark objects was thought to be a straightforward test of the existence of 
massive compact halo objects (MACHOs), which may comprise some of the
dark matter~\citep{paczynski_gravitational_1986}. When the experiment
was carried out, 40\% of the microlensing optical depth was indeed
unexplained by Galactic populations, such as the thick disk and
halo. However, this signal is too small to be caused by MACHOs if they
comprise the entirety of the Milky Way's dark matter halo. Several
authors attempt to explain this excess with stellar populations at
various points along the line of sight to the LMC
\citep{zhao_microlensing_1998,evans_is_2000}, though the viability of
this explanation has also been
disputed~\citep{gould_lmc_1997,gould_new_1999}. \citet{besla_origin_2013}
modelled the interaction of the LMC with the SMC and found that the
microlensing might be explained by clumpy tidal debris from the SMC
being microlensed by the LMC disk. Here, we consider whether our
substantial population of neutron stars and black holes contributes to
the microlensing optical depth to the LMC. We use the formula of
\cite{gould_new_1999} for the required surface mass density $\Sigma$
to contribute lensing optical depth $\tau_{\rm p}$,
\begin{equation}
\Sigma = 47 \left(\frac{\tau_{\mathrm{p}}}{2.9\times10^{-7}}\right) \left(\frac{\hat{D}}{10\;\mathrm{kpc}}\right)^{-1}\;\mathrm{M}_{\odot}\;\mathrm{pc}^{-2}, \qquad
\hat{D} \equiv \frac{d_{\mathrm{ol}} d_{\mathrm{ls}}}{d_{\mathrm{os}}},
\end{equation}
where $d_{\mathrm{ol}}$, $d_{\mathrm{ls}}$ and
$d_{\mathrm{os}}$ are the respective observer-lens, lens-source and
observer-source distances.  We find that our remnants contribute
$0.0035\%$ to the observed microlensing optical depth. In our
calculations, we only include those remnants in front of the LMC and
within three degrees of the sightline between the observer and the
centre of the LMC.

Less familiar than photometric microlensing is the accompanying
astrometric effect, in which the light centroid of the source is
deflected by the presence of the foreground lens.
\citet{belokurov_astrometric_2002} calculated the all-sky photometric
and astrometric microlensing optical depths detectable by \emph{Gaia}
and found that the astrometric optical depth was two orders of
magnitude larger than the photometric optical depth. We calculate the
astrometric optical depth $\tau_{\mathrm{a}}$ for our neutron star and black hole
population using Equation 14 from \citet{belokurov_astrometric_2002},
\begin{equation}
\tau_{\mathrm{a}}=4\sqrt{\frac{\mathrm{G}}{\mathrm{c}^2}}d_{\mathrm{os}}\langle M^{-1/2} \rangle \sqrt{\frac{T_{\mathrm{life}}^3v^3}{5\sqrt{2}\sigma_{\mathrm{a}}}} \int_0^1 \rho(x)\sqrt{1-x}dx,
\end{equation}
where $\mathrm{G}$ is the gravitational constant, $\mathrm{c}$ is the speed of light, $\langle M^{-1/2} \rangle$ is the mean of the inverse square-root of the masses of the compact remnants, $T_{\mathrm{life}}=5\;\mathrm{yr}$ is the estimated lifetime of
\emph{Gaia}, $\sigma_{\mathrm{a}} = 390\;\mu\mathrm{as}$ is the
predicted mean position accuracy of \emph{Gaia} for sources with
$G=18\;\mathrm{mag}$, $v\sim 140\;\mathrm{km}\;\mathrm{s}^{-1}$ is
a characteristic velocity of the lens relative to the LMC disk and $\rho(x)$ is the mass density at a fraction $x$ along the line-of-sight to the source. We
find $\tau_{\mathrm{a}}=1.0\times10^{-10}$ which is 15 times
greater than the corresponding photometric microlensing optical depth.
However, this optical depth is likely still too small to give
observable consequences.

\section{Conclusions}
\label{sec:conclusion}

We have presented a novel source of hypervelocity stars (HVSs) in the
Milky Way (MW) halo. In our model, HVSs originate as runaway stars
from the Large Magellanic Cloud (LMC). The known HVSs possess the
kinematics expected of stars which have been ejected from the LMC, and
thus an LMC origin for some of these stars must be considered a
realistic possibility.

There are a number of current observations that support our scenario,
albeit indirectly.  This includes: (i) a sample of the 31 brightest
stars in the LMC which are consistent with runaway expectations except
perhaps from one anomalously fast supergiant \citep{lennon_gaia_2016},
(ii) young stars in the periphery of the LMC far from star formation
regions \citep{bidin_young_2017} and (iii) B-type stars in the gaseous
leading arm of the LMC with LMC kinematics and chemistry whose
anomalous single nature is in line with a runaway origin
\citep{zhang_chemical_2017}.

The HVSs found in the Sloan Digital Sky Survey footprint are B-type
stars with masses exceeding $3\;\mathrm{M}_{\odot}$. In our model, the
LMC runaways that end up as hypervelocity in the Sloan footprint have
somewhat smaller masses, typically between
$1.6\;\mathrm{M}_{\odot}<M<3.0\;\mathrm{M}_{\odot}$.  However, there
is a strong dependency of the mass and colour of the produced HVSs on
the common-envelope prescription, with lower common-envelope ejection
efficiencies broadly associated with higher mass hypervelocity stars.
So, this discrepancy could be resolved by modest changes to the
uncertain prescription of common envelope evolution.  Alternatively,
the observed HVS population may have contributions from multiple processes only one of which is the fast moving LMC runaway stars.

Our model leads to predictions of the spatial and kinematic signatures
of HVSs seen by \emph{Gaia} and the hypervelocity pulsars observed by the
Square Kilometre Array. We predict that both will be preferentially
found along the past and future orbit of the LMC. The final
\emph{Gaia} catalogue aims to be complete down to
$G\approx20.7$ subject to crowding in dense fields. This
should detect a large number of hypervelocity runaways from the
LMC. We would expect about $200$ of these stars at distances
$30<d<120\;\mathrm{kpc}$ and with proper motions around
$1\;\mathrm{mas}$. This corresponds to a (heliocentric) tangential
velocity of around $500\;\mathrm{km}\;\mathrm{s}^{-1}$ at the location of the LMC. However, we do not expect either parallax or radial
velocities for these stars from \emph{Gaia}, so identification of
their nature will rely on photometric distances and spectroscopy.

In investigating the runaway processes in the LMC, we have linked a
binary stellar evolution code with an N-body model of the interaction
between the Galaxy and the LMC, which enabled us to make powerful
predictions. A problem which required bringing together stellar
evolution and stellar dynamics has implications for both. LMC runaway
stars can provide important constraints on both common-envelope
dispersal and the escape velocity of the Milky Way.

Elsewhere, we have argued that a super-massive black hole (SMBH) in
the LMC may generate HVSs by the Hills mechanism
\citep{boubert_dipole_2016}. This remains plausible, though evidence
for an SMBH in the LMC is elusive at present. However, runaway stars
are a natural consequence of binary evolution in a star-forming
galaxy, and hence they will certainly exist in the LMC. The
exceptionally fast runaways, which become HVSs with respect to the
Milky Way, are sensitive to the prescription of binary
evolution. Changing the binary evolution only seems to modify
the properties of those HVSs and not their number or distribution on the
sky. Our argument therefore does not rely on the precise details of binary
evolution. Furthermore, there are observed counterparts to our
evolutionary channel. A pulsar -- helium white dwarf binary is left
behind if the system is not unbound during the supernova, but is close
enough after the end of common-envelope evolution that the companion
is stripped before igniting helium. The extreme velocity of the
runaways originates in the orbital velocity of such close binaries. We
conclude that hypervelocity runaway stars from the LMC, as a
consequence of star-formation, are unavoidable. They must contribute
to the Galactic HVS population.  The only argument is whether this
process is dominant or subordinate.

\section*{Acknowledgements}
The authors would like to thank the anonymous reviewer whose
suggestion of a plot of apparent magnitude and radial velocity
improved the usefulness of our prediction for observers. The authors thank Mathieu Renzo, Simon
Stevenson, Manos Zapartas and the many other authors cited above who
have contributed to the development of {\sc binary\_c}. We also
thank Vasily Belokurov and other members of the Streams discussion
group for comments on this work as it was under development. DB thanks
Isaac Shivers for a thought-provoking visualisation of the Galactic
pulsar distribution
(\url{http://w.astro.berkeley.edu/~ishivvers/pulsars.html}).  DB is
grateful to the Science and Technology Facilities Council (STFC) for
providing PhD funding. DE acknowledges that research leading to these results has received
funding from the European Research Council under the European Union's
Seventh Framework Programme (FP/2007-2013)/ERC Grant Agreement
no. 308024. RGI thanks the STFC for funding his Rutherford fellowship under grant ST/L003910/1 and Churchill College, Cambridge for his fellowship.




\bibliographystyle{mnras}
\bibliography{references} 

\begin{thebibliography}{}
\makeatletter
\relax
\def\mn@urlcharsother{\let\do\@makeother \do\$\do\&\do\#\do\^\do\_\do\%\do\~}
\def\mn@doi{\begingroup\mn@urlcharsother \@ifnextchar [ {\mn@doi@}
  {\mn@doi@[]}}
\def\mn@doi@[#1]#2{\def\@tempa{#1}\ifx\@tempa\@empty \href
  {http://dx.doi.org/#2} {doi:#2}\else \href {http://dx.doi.org/#2} {#1}\fi
  \endgroup}
\def\mn@eprint#1#2{\mn@eprint@#1:#2::\@nil}
\def\mn@eprint@arXiv#1{\href {http://arxiv.org/abs/#1} {{\tt arXiv:#1}}}
\def\mn@eprint@dblp#1{\href {http://dblp.uni-trier.de/rec/bibtex/#1.xml}
  {dblp:#1}}
\def\mn@eprint@#1:#2:#3:#4\@nil{\def\@tempa {#1}\def\@tempb {#2}\def\@tempc
  {#3}\ifx \@tempc \@empty \let \@tempc \@tempb \let \@tempb \@tempa \fi \ifx
  \@tempb \@empty \def\@tempb {arXiv}\fi \@ifundefined
  {mn@eprint@\@tempb}{\@tempb:\@tempc}{\expandafter \expandafter \csname
  mn@eprint@\@tempb\endcsname \expandafter{\@tempc}}}

\bibitem[\protect\citeauthoryear{Abadi, Navarro  \& Steinmetz}{Abadi
  et~al.}{2009}]{abadi_alternative_2009}
Abadi M.~G.,  Navarro J.~F.,   Steinmetz M.,  2009, \mn@doi [ApJL]
  {10.1088/0004-637X/691/2/L63}, 691, L63

\bibitem[\protect\citeauthoryear{Abate, Pols, Izzard, Mohamed  \& de
  Mink}{Abate et~al.}{2013}]{abate_wind_2013}
Abate C.,  Pols O.~R.,  Izzard R.~G.,  Mohamed S.~S.,   de Mink S.~E.,  2013,
  \mn@doi [A\&A] {10.1051/0004-6361/201220007}, 552, A26

\bibitem[\protect\citeauthoryear{Abate, Pols, Stancliffe, Izzard, Karakas,
  Beers  \& Lee}{Abate et~al.}{2015}]{abate_modelling_2015}
Abate C.,  Pols O.~R.,  Stancliffe R.~J.,  Izzard R.~G.,  Karakas A.~I.,  Beers
  T.~C.,   Lee Y.~S.,  2015, \mn@doi [A\&A] {10.1051/0004-6361/201526200}, 581,
  A62

\bibitem[\protect\citeauthoryear{Althaus \& Benvenuto}{Althaus \&
  Benvenuto}{1997}]{althaus_evolution_1997}
Althaus L.~G.,  Benvenuto O.~G.,  1997, \mn@doi [ApJ] {10.1086/303686}, 477,
  313

\bibitem[\protect\citeauthoryear{Arenou}{Arenou}{2010}]{arenou_simulated_2010}
Arenou F.,  2010, The simulated multiple stars, \url
  {www.rssd.esa.int/doc_fetch.php?id=2969346}

\bibitem[\protect\citeauthoryear{Backer}{Backer}{1998}]{backer_neutron_1998}
Backer D.~C.,  1998, \mn@doi [ApJ] {10.1086/305167}, 493, 873

\bibitem[\protect\citeauthoryear{Belokurov \& Evans}{Belokurov \&
  Evans}{2002}]{belokurov_astrometric_2002}
Belokurov V.~A.,  Evans N.~W.,  2002, \mn@doi [MNRAS]
  {10.1046/j.1365-8711.2002.05222.x}, 331, 649

\bibitem[\protect\citeauthoryear{Besla, Hernquist  \& Loeb}{Besla
  et~al.}{2013}]{besla_origin_2013}
Besla G.,  Hernquist L.,   Loeb A.,  2013, \mn@doi [MNRAS]
  {10.1093/mnras/sts192}, 428, 2342

\bibitem[\protect\citeauthoryear{Bidin, Casetti-Dinescu, Girard, Zhang, Mendez,
  Vieira, Korchagin  \& van Altena}{Bidin et~al.}{2017}]{bidin_young_2017}
Bidin C.~M.,  Casetti-Dinescu D.~I.,  Girard T.~M.,  Zhang L.,  Mendez R.~A.,
  Vieira K.,  Korchagin V.~I.,   van Altena W.~F.,  2017, \mn@doi [MNRAS]
  {10.1093/mnras/stw3242}, 466, 3077

\bibitem[\protect\citeauthoryear{Blaauw}{Blaauw}{1961}]{blaauw_origin_1961}
Blaauw A.,  1961, Bulletin of the Astronomical Institutes of the Netherlands,
  15, 265

\bibitem[\protect\citeauthoryear{Boubert \& Evans}{Boubert \&
  Evans}{2016}]{boubert_dipole_2016}
Boubert D.,  Evans N.~W.,  2016, \mn@doi [ApJL] {10.3847/2041-8205/825/1/L6},
  825, L6

\bibitem[\protect\citeauthoryear{Bromley, Kenyon, Brown  \& Geller}{Bromley
  et~al.}{2009}]{bromley_runaway_2009}
Bromley B.~C.,  Kenyon S.~J.,  Brown W.~R.,   Geller M.~J.,  2009, \mn@doi
  [ApJ] {10.1088/0004-637X/706/2/925}, 706, 925

\bibitem[\protect\citeauthoryear{Brown}{Brown}{2015}]{brown_hypervelocity_2015}
Brown W.~R.,  2015, \mn@doi [ARA\&A] {10.1146/annurev-astro-082214-122230}, 53,
  15

\bibitem[\protect\citeauthoryear{Brown, Geller, Kenyon  \& Kurtz}{Brown
  et~al.}{2005}]{brown_discovery_2005}
Brown W.~R.,  Geller M.~J.,  Kenyon S.~J.,   Kurtz M.~J.,  2005, \mn@doi [ApJL]
  {10.1086/429378}, 622, L33

\bibitem[\protect\citeauthoryear{Brown, Geller, Kenyon  \& Kurtz}{Brown
  et~al.}{2006a}]{brown_successful_2006}
Brown W.~R.,  Geller M.~J.,  Kenyon S.~J.,   Kurtz M.~J.,  2006a, \mn@doi
  [ApJL] {10.1086/503279}, 640, L35

\bibitem[\protect\citeauthoryear{Brown, Geller, Kenyon  \& Kurtz}{Brown
  et~al.}{2006b}]{brown_hypervelocity_2006}
Brown W.~R.,  Geller M.~J.,  Kenyon S.~J.,   Kurtz M.~J.,  2006b, \mn@doi [ApJ]
  {10.1086/505165}, 647, 303

\bibitem[\protect\citeauthoryear{Brown, Geller, Kenyon, Kurtz  \&
  Bromley}{Brown et~al.}{2007a}]{brown_hypervelocity_2007-1}
Brown W.~R.,  Geller M.~J.,  Kenyon S.~J.,  Kurtz M.~J.,   Bromley B.~C.,
  2007a, \mn@doi [ApJ] {10.1086/513595}, 660, 311

\bibitem[\protect\citeauthoryear{Brown, Geller, Kenyon, Kurtz  \&
  Bromley}{Brown et~al.}{2007b}]{brown_hypervelocity_2007}
Brown W.~R.,  Geller M.~J.,  Kenyon S.~J.,  Kurtz M.~J.,   Bromley B.~C.,
  2007b, \mn@doi [ApJ] {10.1086/523642}, 671, 1708

\bibitem[\protect\citeauthoryear{Brown, Geller, Kenyon  \& Bromley}{Brown
  et~al.}{2009a}]{brown_anisotropic_2009}
Brown W.~R.,  Geller M.~J.,  Kenyon S.~J.,   Bromley B.~C.,  2009a, \mn@doi
  [ApJL] {10.1088/0004-637X/690/1/L69}, 690, L69

\bibitem[\protect\citeauthoryear{Brown, Geller  \& Kenyon}{Brown
  et~al.}{2009b}]{brown_mmt_2009}
Brown W.~R.,  Geller M.~J.,   Kenyon S.~J.,  2009b, \mn@doi [ApJ]
  {10.1088/0004-637X/690/2/1639}, 690, 1639

\bibitem[\protect\citeauthoryear{Brown, Geller  \& Kenyon}{Brown
  et~al.}{2012}]{brown_mmt_2012}
Brown W.~R.,  Geller M.~J.,   Kenyon S.~J.,  2012, \mn@doi [ApJ]
  {10.1088/0004-637X/751/1/55}, 751, 55

\bibitem[\protect\citeauthoryear{Brown, Geller  \& Kenyon}{Brown
  et~al.}{2014}]{brown_mmt_2014}
Brown W.~R.,  Geller M.~J.,   Kenyon S.~J.,  2014, \mn@doi [ApJ]
  {10.1088/0004-637X/787/1/89}, 787, 89

\bibitem[\protect\citeauthoryear{Brown, Anderson, Gnedin, Bond, Geller  \&
  Kenyon}{Brown et~al.}{2015}]{brown_proper_2015}
Brown W.~R.,  Anderson J.,  Gnedin O.~Y.,  Bond H.~E.,  Geller M.~J.,   Kenyon
  S.~J.,  2015, \mn@doi [ApJ] {10.1088/0004-637X/804/1/49}, 804, 49

\bibitem[\protect\citeauthoryear{Casetti-Dinescu, Vieira, Girard  \& van
  Altena}{Casetti-Dinescu et~al.}{2012}]{casetti-dinescu_constraints_2012}
Casetti-Dinescu D.~I.,  Vieira K.,  Girard T.~M.,   van Altena W.~F.,  2012,
  \mn@doi [ApJ] {10.1088/0004-637X/753/2/123}, 753, 123

\bibitem[\protect\citeauthoryear{Casetti-Dinescu, Bidin, Girard, Mendez,
  Vieira, Korchagin  \& van Altena}{Casetti-Dinescu
  et~al.}{2014}]{casetti-dinescu_recent_2014}
Casetti-Dinescu D.~I.,  Bidin C.~M.,  Girard T.~M.,  Mendez R.~A.,  Vieira K.,
  Korchagin V.~I.,   van Altena W.~F.,  2014, \mn@doi [ApJ]
  {10.1088/2041-8205/784/2/L37}, 784, L37

\bibitem[\protect\citeauthoryear{Cignoni et~al.,}{Cignoni
  et~al.}{2015}]{cignoni_hubble_2015}
Cignoni M.,  et~al., 2015, \mn@doi [ApJ] {10.1088/0004-637X/811/2/76}, 811, 76

\bibitem[\protect\citeauthoryear{Claeys, Pols, Izzard, Vink  \& Verbunt}{Claeys
  et~al.}{2014}]{claeys_theoretical_2014}
Claeys J. S.~W.,  Pols O.~R.,  Izzard R.~G.,  Vink J.,   Verbunt F. W.~M.,
  2014, \mn@doi [A\&A] {10.1051/0004-6361/201322714}, 563, A83

\bibitem[\protect\citeauthoryear{Cordes \& Chernoff}{Cordes \&
  Chernoff}{1998}]{cordes_neutron_1998}
Cordes J.~M.,  Chernoff D.~F.,  1998, \mn@doi [ApJ] {10.1086/306138}, 505, 315

\bibitem[\protect\citeauthoryear{De~Marchi et~al.,}{De~Marchi
  et~al.}{2011}]{de_marchi_star_2011}
De~Marchi G.,  et~al., 2011, \mn@doi [ApJ] {10.1088/0004-637X/739/1/27}, 739,
  27

\bibitem[\protect\citeauthoryear{De~Marco \& Izzard}{De~Marco \&
  Izzard}{2017}]{de_marco_dawes_2017}
De~Marco O.,  Izzard R.~G.,  2017, \mn@doi [PASA] {10.1017/pasa.2016.52}, 34,
  e001

\bibitem[\protect\citeauthoryear{de Mink, Brott, Cantiello, Izzard, Langer  \&
  Sana}{de~Mink et~al.}{2012}]{de_mink_challenges_2012}
de Mink S.~E.,  Brott I.,  Cantiello M.,  Izzard R.~G.,  Langer N.,   Sana H.,
  2012. Challenges for Understanding the Evolution of Massive Stars: Rotation,
  Binarity, and Mergers, Proceedings of a Scientific Meeting in Honor of
  Anthony F. J. Moffat, p.~65

\bibitem[\protect\citeauthoryear{de Mink, Langer, Izzard, Sana  \& de
  Koter}{de~Mink et~al.}{2013}]{de_mink_rotation_2013}
de Mink S.~E.,  Langer N.,  Izzard R.~G.,  Sana H.,   de Koter A.,  2013,
  \mn@doi [ApJ] {10.1088/0004-637X/764/2/166}, 764, 166

\bibitem[\protect\citeauthoryear{de Mink, Sana, Langer, Izzard  \&
  Schneider}{de~Mink et~al.}{2014}]{de_mink_incidence_2014}
de Mink S.~E.,  Sana H.,  Langer N.,  Izzard R.~G.,   Schneider F. R.~N.,
  2014, \mn@doi [ApJ] {10.1088/0004-637X/782/1/7}, 782, 7

\bibitem[\protect\citeauthoryear{Dewi \& Tauris}{Dewi \&
  Tauris}{2000}]{dewi_energy_2000}
Dewi J. D.~M.,  Tauris T.~M.,  2000, A\&A, 360, 1043

\bibitem[\protect\citeauthoryear{Duquennoy \& Mayor}{Duquennoy \&
  Mayor}{1991}]{duquennoy_multiplicity_1991}
Duquennoy A.,  Mayor M.,  1991, A\&A, 248, 485

\bibitem[\protect\citeauthoryear{Edelmann, Napiwotzki, Heber, Christlieb  \&
  Reimers}{Edelmann et~al.}{2005}]{edelmann_he_2005}
Edelmann H.,  Napiwotzki R.,  Heber U.,  Christlieb N.,   Reimers D.,  2005,
  \mn@doi [ApJL] {10.1086/498940}, 634, L181

\bibitem[\protect\citeauthoryear{Evans \& Kerins}{Evans \&
  Kerins}{2000}]{evans_is_2000}
Evans N.~W.,  Kerins E.,  2000, \mn@doi [ApJ] {10.1086/308328}, 529, 917

\bibitem[\protect\citeauthoryear{Evans \& Massey}{Evans \&
  Massey}{2015}]{evans_runaway_2015}
Evans K.~A.,  Massey P.,  2015, \mn@doi [AJ] {10.1088/0004-6256/150/5/149},
  150, 149

\bibitem[\protect\citeauthoryear{Fryer, Burrows  \& Benz}{Fryer
  et~al.}{1998}]{fryer_population_1998}
Fryer C.,  Burrows A.,   Benz W.,  1998, \mn@doi [ApJ] {10.1086/305348}, 496,
  333

\bibitem[\protect\citeauthoryear{Geier et~al.,}{Geier
  et~al.}{2013}]{geier_progenitor_2013}
Geier S.,  et~al., 2013, \mn@doi [A\&A] {10.1051/0004-6361/201321395}, 554, A54

\bibitem[\protect\citeauthoryear{Gould}{Gould}{1997}]{gould_lmc_1997}
Gould A.,  1997, preprint (arXiv:astro-ph/9709263)

\bibitem[\protect\citeauthoryear{Gould}{Gould}{1999}]{gould_new_1999}
Gould A.,  1999, \mn@doi [ApJ] {10.1086/307930}, 525, 734

\bibitem[\protect\citeauthoryear{Gualandris \& Portegies~Zwart}{Gualandris \&
  Portegies~Zwart}{2007}]{gualandris_hypervelocity_2007}
Gualandris A.,  Portegies~Zwart S.,  2007, \mn@doi [MNRAS]
  {10.1111/j.1745-3933.2007.00280.x}, 376, L29

\bibitem[\protect\citeauthoryear{Hansen \& Phinney}{Hansen \&
  Phinney}{1997}]{hansen_pulsar_1997}
Hansen B. M.~S.,  Phinney E.~S.,  1997, \mn@doi [MNRAS]
  {10.1093/mnras/291.3.569}, 291, 569

\bibitem[\protect\citeauthoryear{Harris \& Zaritsky}{Harris \&
  Zaritsky}{2009}]{harris_star_2009}
Harris J.,  Zaritsky D.,  2009, \mn@doi [AJ] {10.1088/0004-6256/138/5/1243},
  138, 1243

\bibitem[\protect\citeauthoryear{Hills}{Hills}{1988}]{hills_hyper-velocity_1988}
Hills J.~G.,  1988, \mn@doi [Nature] {10.1038/331687a0}, 331, 687

\bibitem[\protect\citeauthoryear{Hirsch, Heber, O'Toole  \& Bresolin}{Hirsch
  et~al.}{2005}]{hirsch_us_2005}
Hirsch H.~A.,  Heber U.,  O'Toole S.~J.,   Bresolin F.,  2005, \mn@doi [A\&A]
  {10.1051/0004-6361:200500212}, 444, L61

\bibitem[\protect\citeauthoryear{Hurley, Tout  \& Pols}{Hurley
  et~al.}{2002}]{hurley_evolution_2002}
Hurley J.~R.,  Tout C.~A.,   Pols O.~R.,  2002, \mn@doi [MNRAS]
  {10.1046/j.1365-8711.2002.05038.x}, 329, 897

\bibitem[\protect\citeauthoryear{{Kallivayalil}, {van der Marel}, {Besla},
  {Anderson}  \& {Alcock}}{{Kallivayalil}
  et~al.}{2013}]{kallivayalil_etal_2013}
{Kallivayalil} N.,  {van der Marel} R.~P.,  {Besla} G.,  {Anderson} J.,
  {Alcock} C.,  2013, \mn@doi [\apj] {10.1088/0004-637X/764/2/161}, \href
  {http://adsabs.harvard.edu/abs/2013ApJ...764..161K} {764, 161}

\bibitem[\protect\citeauthoryear{Izzard, Tout, Karakas  \& Pols}{Izzard
  et~al.}{2004}]{izzard_new_2004}
Izzard R.~G.,  Tout C.~A.,  Karakas A.~I.,   Pols O.~R.,  2004, \mn@doi [MNRAS]
  {10.1111/j.1365-2966.2004.07446.x}, 350, 407

\bibitem[\protect\citeauthoryear{Izzard, Dray, Karakas, Lugaro  \& Tout}{Izzard
  et~al.}{2006}]{izzard_population_2006}
Izzard R.~G.,  Dray L.~M.,  Karakas A.~I.,  Lugaro M.,   Tout C.~A.,  2006,
  \mn@doi [A\&A] {10.1051/0004-6361:20066129}, 460, 565

\bibitem[\protect\citeauthoryear{Izzard, Glebbeek, Stancliffe  \& Pols}{Izzard
  et~al.}{2009}]{izzard_population_2009}
Izzard R.~G.,  Glebbeek E.,  Stancliffe R.~J.,   Pols O.~R.,  2009, \mn@doi
  [A\&A] {10.1051/0004-6361/200912827}, 508, 1359

\bibitem[\protect\citeauthoryear{Jethwa, Erkal  \& Belokurov}{Jethwa
  et~al.}{2016}]{jethwa_magellanic_2016}
Jethwa P.,  Erkal D.,   Belokurov V.,  2016, \mn@doi [MNRAS]
  {10.1093/mnras/stw1343}, 461, 2212

\bibitem[\protect\citeauthoryear{Jordi et~al.,}{Jordi
  et~al.}{2010}]{jordi_gaia_2010}
Jordi C.,  et~al., 2010, \mn@doi [A\&A] {10.1051/0004-6361/201015441}, 523, A48

\bibitem[\protect\citeauthoryear{Justham, Wolf, Podsiadlowski  \& Han}{Justham
  et~al.}{2009}]{justham_type_2009}
Justham S.,  Wolf C.,  Podsiadlowski P.,   Han Z.,  2009, \mn@doi [A\&A] {10.1051/0004-6361:200810106}, 493, 1081

\bibitem[\protect\citeauthoryear{Kenyon, Bromley, Brown  \& Geller}{Kenyon
  et~al.}{2014}]{kenyon_predicted_2014}
Kenyon S.~J.,  Bromley B.~C.,  Brown W.~R.,   Geller M.~J.,  2014, \mn@doi
  [ApJ] {10.1088/0004-637X/793/2/122}, 793, 122

\bibitem[\protect\citeauthoryear{Kroupa}{Kroupa}{2001}]{kroupa_variation_2001}
Kroupa P.,  2001, \mn@doi [MNRAS] {10.1046/j.1365-8711.2001.04022.x}, 322, 231

\bibitem[\protect\citeauthoryear{Lada \& Lada}{Lada \&
  Lada}{2003}]{lada_embedded_2003}
Lada C.~J.,  Lada E.~A.,  2003, \mn@doi [ARA\&A]
  {10.1146/annurev.astro.41.011802.094844}, 41, 57

\bibitem[\protect\citeauthoryear{Lennon, van~der Marel, Lerate, O'Mullane  \&
  Sahlmann}{Lennon et~al.}{2016}]{lennon_gaia_2016}
Lennon D.~J.,  van~der Marel R.~P.,  Lerate M.~R.,  O'Mullane W.,   Sahlmann
  J.,  2016, preprint (arXiv:1611.05504)

\bibitem[\protect\citeauthoryear{Liu, Tauris, R{\"o}pke, Moriya, Kruckow,
  Stancliffe  \& Izzard}{Liu et~al.}{2015}]{liu_interaction_2015}
Liu Z.-W.,  Tauris T.~M.,  R{\"o}pke F.~K.,  Moriya T.~J.,  Kruckow M.,
  Stancliffe R.~J.,   Izzard R.~G.,  2015, \mn@doi [A\&A]
  {10.1051/0004-6361/201526757}, 584, A11

\bibitem[\protect\citeauthoryear{Mackey, Koposov, Erkal, Belokurov, Da~Costa
  \& Gomez}{Mackey et~al.}{2016}]{mackey_10_2016}
Mackey A.~D.,  Koposov S.~E.,  Erkal D.,  Belokurov V.,  Da~Costa G.~S.,
  Gomez F.~A.,  2016, \mn@doi [MNRAS] {10.1093/mnras/stw497}, 459, 239

\bibitem[\protect\citeauthoryear{Manchester, Hobbs, Teoh  \& Hobbs}{Manchester
  et~al.}{2005}]{manchester_australia_2005}
Manchester R.~N.,  Hobbs G.~B.,  Teoh A.,   Hobbs M.,  2005, \mn@doi [AJ]
  {10.1086/428488}, 129, 1993

\bibitem[\protect\citeauthoryear{Martin, Jeffery, Naslim  \& Woolf}{Martin
  et~al.}{2017}]{martin_kinematics_2017}
Martin P.,  Jeffery C.~S.,  Naslim N.,   Woolf V.~M.,  2017, preprint (arXiv:1701.03026)
  

\bibitem[\protect\citeauthoryear{Opik}{Opik}{1924}]{opik_statistical_1924}
Opik E.,  1924, Publications of the Tartu Astrofizica Observatory, 25

\bibitem[\protect\citeauthoryear{Paczynski}{Paczynski}{1986}]{paczynski_gravitational_1986}
Paczynski B.,  1986, \mn@doi [ApJ] {10.1086/164140}, 304, 1

\bibitem[\protect\citeauthoryear{{Pe{\~n}arrubia}, {G{\'o}mez}, {Besla},
  {Erkal}  \& {Ma}}{{Pe{\~n}arrubia} et~al.}{2016}]{penarrubia_etal_2016}
{Pe{\~n}arrubia} J.,  {G{\'o}mez} F.~A.,  {Besla} G.,  {Erkal} D.,   {Ma}
  Y.-Z.,  2016, \mn@doi [\mnras] {10.1093/mnrasl/slv160}, \href
  {http://adsabs.harvard.edu/abs/2016MNRAS.456L..54P} {456, L54}

\bibitem[\protect\citeauthoryear{Perets}{Perets}{2009}]{perets_dynamical_2009}
Perets H.~B.,  2009, \mn@doi [ApJ] {10.1088/0004-637X/690/1/795}, 690, 795

\bibitem[\protect\citeauthoryear{Perets \& {\v S}ubr}{Perets \&
  {\v S}ubr}{2012}]{perets_properties_2012}
Perets H.~B.,  {\v S}ubr L.,  2012, \mn@doi [ApJ] {10.1088/0004-637X/751/2/133},
  751, 133

\bibitem[\protect\citeauthoryear{Piatti \& Geisler}{Piatti \&
  Geisler}{2013}]{piatti_age-metallicity_2013}
Piatti A.~E.,  Geisler D.,  2013, \mn@doi [AJ] {10.1088/0004-6256/145/1/17},
  145, 17

\bibitem[\protect\citeauthoryear{Podsiadlowski, Pfahl  \&
  Rappaport}{Podsiadlowski et~al.}{2005}]{podsiadlowski_neutron-star_2005}
Podsiadlowski P.,  Pfahl E.,   Rappaport S.,  2005. Binary Radio Pulsars,
  Astronomical Society of the Pacific Conference Series, p.~327

\bibitem[\protect\citeauthoryear{Portegies~Zwart}{Portegies~Zwart}{2000}]{portegies_zwart_characteristics_2000}
Portegies~Zwart S.~F.,  2000, \mn@doi [ApJ] {10.1086/317190}, 544, 437

\bibitem[\protect\citeauthoryear{Przybilla, Nieva, Heber, Firnstein, Butler,
  Napiwotzki  \& Edelmann}{Przybilla et~al.}{2008}]{przybilla_lmc_2008}
Przybilla N.,  Nieva M.~F.,  Heber U.,  Firnstein M.,  Butler K.,  Napiwotzki
  R.,   Edelmann H.,  2008, \mn@doi [A\&A] {10.1051/0004-6361:200809391}, 480,
  L37

\bibitem[\protect\citeauthoryear{Raghavan et~al.,}{Raghavan
  et~al.}{2010}]{raghavan_survey_2010}
Raghavan D.,  et~al., 2010, \mn@doi [ApJS] {10.1088/0067-0049/190/1/1}, 190, 1

\bibitem[\protect\citeauthoryear{Ridley \& Lorimer}{Ridley \&
  Lorimer}{2010}]{ridley_new_2010}
Ridley J.~P.,  Lorimer D.~R.,  2010, \mn@doi [MNRAS] {10.1111/j.1745-3933.2010.00886.x}, 406, L80

\bibitem[\protect\citeauthoryear{Sana et~al.,}{Sana
  et~al.}{2012}]{sana_binary_2012}
Sana H.,  et~al., 2012, \mn@doi [Science] {10.1126/science.1223344}, 337, 444

\bibitem[\protect\citeauthoryear{Schlegel, Finkbeiner  \& Davis}{Schlegel
  et~al.}{1998}]{schlegel_maps_1998}
Schlegel D.~J.,  Finkbeiner D.~P.,   Davis M.,  1998, \mn@doi [ApJ]
  {10.1086/305772}, 500, 525

\bibitem[\protect\citeauthoryear{Schneider et~al.,}{Schneider
  et~al.}{2014}]{schneider_ages_2014}
Schneider F. R.~N.,  et~al., 2014, \mn@doi [ApJ] {10.1088/0004-637X/780/2/117},
  780, 117

\bibitem[\protect\citeauthoryear{Sch{\"o}nrich, Binney  \& Dehnen}{Sch{\"o}nrich
  et~al.}{2010}]{schonrich_local_2010}
Sch{\"o}nrich R.,  Binney J.,   Dehnen W.,  2010, \mn@doi [MNRAS]
  {10.1111/j.1365-2966.2010.16253.x}, 403, 1829

\bibitem[\protect\citeauthoryear{Smith \& Tombleson}{Smith \&
  Tombleson}{2015}]{smith_luminous_2015}
Smith N.,  Tombleson R.,  2015, \mn@doi [MNRAS] {10.1093/mnras/stu2430}, 447,
  598

\bibitem[\protect\citeauthoryear{Smits, Kramer, Stappers, Lorimer, Cordes  \&
  Faulkner}{Smits et~al.}{2009}]{smits_pulsar_2009}
Smits R.,  Kramer M.,  Stappers B.,  Lorimer D.~R.,  Cordes J.,   Faulkner A.,
  2009, \mn@doi [A\&A] {10.1051/0004-6361:200810383}, 493, 1161

\bibitem[\protect\citeauthoryear{Spera, Mapelli  \& Bressan}{Spera
  et~al.}{2015}]{spera_mass_2015}
Spera M.,  Mapelli M.,   Bressan A.,  2015, \mn@doi [MNRAS]
  {10.1093/mnras/stv1161}, 451, 4086

\bibitem[\protect\citeauthoryear{Springel}{Springel}{2005}]{springel_cosmological_2005}
Springel V.,  2005, \mn@doi [MNRAS] {10.1111/j.1365-2966.2005.09655.x}, 364,
  1105

\bibitem[\protect\citeauthoryear{Tauris \& Dewi}{Tauris \&
  Dewi}{2001}]{tauris_binding_2001}
Tauris T.~M.,  Dewi J. D.~M.,  2001, \mn@doi [A\&A]
  {10.1051/0004-6361:20010099}, 369, 170

\bibitem[\protect\citeauthoryear{Tauris \& Takens}{Tauris \&
  Takens}{1998}]{tauris_runaway_1998}
Tauris T.~M.,  Takens R.~J.,  1998, A\&A, 330, 1047

\bibitem[\protect\citeauthoryear{van~der Marel \& Kallivayalil}{van~der Marel
  \& Kallivayalil}{2014}]{van_der_marel_third-epoch_2014}
van~der Marel R.~P.,  Kallivayalil N.,  2014, \mn@doi [ApJ]
  {10.1088/0004-637X/781/2/121}, 781, 121

\bibitem[\protect\citeauthoryear{Yao, Manchester  \& Wang}{Yao
  et~al.}{2017}]{yao_new_2017}
Yao J.~M.,  Manchester R.~N.,   Wang N.,  2017, \mn@doi [ApJ]
  {10.3847/1538-4357/835/1/29}, 835, 29

\bibitem[\protect\citeauthoryear{Zapartas et~al.,}{Zapartas
  et~al.}{2017}]{zapartas_delay-time_2017}
Zapartas E.,  et~al., 2017, preprint (arXiv:1701.07032)

\bibitem[\protect\citeauthoryear{Zhang et~al.,}{Zhang
  et~al.}{2017}]{zhang_chemical_2017}
Zhang L.,  et~al., 2017, preprint (arXiv:1701.00590)

\bibitem[\protect\citeauthoryear{Zhao}{Zhao}{1998}]{zhao_microlensing_1998}
Zhao H.,  1998, \mn@doi [MNRAS] {10.1046/j.1365-8711.1998.01180.x}, 294, 139

\makeatother
\end{thebibliography}




%
%


\bsp	
\label{lastpage}
\end{document}